\documentclass[10pt,journal,compsoc]{IEEEtran}

\usepackage{cite}
\usepackage{times}
\usepackage{graphicx}
\usepackage{amsthm,amssymb,mathtools}
\usepackage{subfigure}
\usepackage{url}
\usepackage{CJK}
\graphicspath{{figure/}}
\usepackage{multirow}
\usepackage{booktabs}
\usepackage{array}
\usepackage{color}
\usepackage{algorithm}
\usepackage{algorithmic}
\usepackage{enumitem}
\usepackage{soul}
\usepackage{nicefrac}

\newcolumntype{L}[1]{>{\raggedright\let\newline\\\arraybackslash\hspace{0pt}}m{#1}}
\newcolumntype{C}[1]{>{\centering\let\newline  \\\arraybackslash\hspace{0pt}}m{#1}}
\newcolumntype{R}[1]{>{\raggedleft\let\newline \\\arraybackslash\hspace{0pt}}m{#1}}

\newtheorem{assumption}{Assumption}
\newtheorem{theorem}{Theorem}[section]
\newtheorem{lemma}[theorem]{Lemma}
\newtheorem{proposition}[theorem]{Proposition}

\newtheorem{remark}{Remark}[section]

\newcommand{\Px}[2]{\text{prox}_{#1}(#2)}

\begin{document}
%
% paper title
% Titles are generally capitalized except for words such as a, an, and, as,
% at, but, by, for, in, nor, of, on, or, the, to and up, which are usually
% not capitalized unless they are the first or last word of the title.
% Linebreaks \\ can be used within to get better formatting as desired.
% Do not put math or special symbols in the title.
\title{Non-local Meets Global: An Iterative Paradigm for Hyperspectral Image Restoration}
%
%
% author names and IEEE memberships
% note positions of commas and nonbreaking spaces ( ~ ) LaTeX will not break
% a structure at a ~ so this keeps an author's name from being broken across
% two lines.
% use \thanks{} to gain access to the first footnote area
% a separate \thanks must be used for each paragraph as LaTeX2e's \thanks
% was not built to handle multiple paragraphs
%
%
%\IEEEcompsocitemizethanks is a special \thanks that produces the bulleted
% lists the Computer Society journals use for "first footnote" author
% affiliations. Use \IEEEcompsocthanksitem which works much like \item
% for each affiliation group. When not in compsoc mode,
% \IEEEcompsocitemizethanks becomes like \thanks and
% \IEEEcompsocthanksitem becomes a line break with idention. This
% facilitates dual compilation, although admittedly the differences in the
% desired content of \author between the different types of papers makes a
% one-size-fits-all approach a daunting prospect. For instance, compsoc
% journal papers have the author affiliations above the "Manuscript
% received ..."  text while in non-compsoc journals this is reversed. Sigh.

\author{Wei~He,~\IEEEmembership{Member,~IEEE,}
        Quanming~Yao,~\IEEEmembership{Member,~IEEE,}
        Chao~Li,
        Naoto~Yokoya,~\IEEEmembership{Member,~IEEE,}
        Qibin~Zhao,~\IEEEmembership{Senior~Member,~IEEE,} % <-this % stops a space
        Hongyan~Zhang,~\IEEEmembership{Senior~Member,~IEEE,}% <-this % stops a space
        ~Liangpei~Zhang,~\IEEEmembership{Fellow,~IEEE}% <-this % stops a space

%\thanks{This work was supported by the Japan Society for the Promotion of Science (KAKENHI 19K20308 and KAKENHI 18K18067).}
\IEEEcompsocitemizethanks{\IEEEcompsocthanksitem W. He, C. Li, N. Yokoya and Q. Zhao were with RIKEN Center for Advanced Intelligence Project (AIP) Tokyo, 103-0027, Japan.
E-mail: \{wei.he;chao.li;naoto.yokoya;qibin.Zhao\}@riken.jp
\IEEEcompsocthanksitem Q. Yao was with Department of Electronic Engineering,
Tsinghua University and 4Paradigm Inc. Beijing, China.
E-mail: qyaoaa@connect.ust.hk
\IEEEcompsocthanksitem N. Yokoya was also with Graduate School of Frontier Sciences, The University of Tokyo, Chiba, 277-8561, Japan.
\IEEEcompsocthanksitem H. Zhang and L. Zhang were with the LISMARS, Wuhan University, Wuhan, 430072, China.
E-mail: \{zhanghongyan;zlp62\}@whu.edu.cn
}% <-this % stops an unwanted space
\thanks{Manuscript received Sep. 10, 2019; revised Mar. 03, 2020 and Jun. 28, 2020; accepted Sep. 17, 2020. (Corresponding author: Quanming Yao.)}}

% The paper headers
\markboth{Journal of \LaTeX\ Class Files,~Vol.~XX, No.~XX, XXXX~XXXX}%
{Shell \MakeLowercase{\textit{et al.}}: Non-local Meets Global: An Integrated Paradigm for Hyperspectral Restoration}

% abstract or keywords.
\IEEEtitleabstractindextext{%
\begin{abstract}
Non-local low-rank tensor approximation has been developed as a state-of-the-art method for hyperspectral image
(HSI) restoration, which includes the tasks of denoising, compressed HSI reconstruction and inpainting.
Unfortunately, while its restoration performance benefits from more spectral bands,
its runtime also substantially increases.
In this paper, we claim that the HSI lies in a global spectral low-rank subspace, and the spectral subspaces of each full band patch group should lie in this global low-rank subspace.
This motivates us to propose a unified paradigm combining the spatial and spectral properties for HSI restoration.
The proposed paradigm enjoys performance superiority from the non-local spatial denoising and light computation complexity from the low-rank orthogonal basis exploration. An efficient alternating minimization algorithm with rank adaptation is developed.
It is done by first solving a fidelity term-related problem for the update of a latent input image, and then learning a low-dimensional orthogonal basis and the related reduced image from the latent input image.
Subsequently, non-local low-rank denoising is developed to refine the reduced image and orthogonal basis iteratively. Finally, the experiments on HSI denoising, compressed reconstruction, and inpainting tasks, with both simulated and real datasets, demonstrate its superiority with respect to state-of-the-art HSI restoration methods.
\end{abstract}

% Note that keywords are not normally used for peerreview papers.
}

% make the title area
\maketitle
\IEEEdisplaynontitleabstractindextext
\IEEEpeerreviewmaketitle

\IEEEraisesectionheading{\section{Introduction}\label{sec:introduction}}
\IEEEPARstart{H}{yperspectral} imagery (HSI) is a three-dimensional (3D) cube covering the spectral wavelength region from 0.4 to 2.5 $\mu m$ at a nominal spectral resolution of less than 10 nm~\cite{Green1998}. Thanks to the fast development of HSI techniques~\cite{kwon2005kernel,HarvardCVPR2011,CAVETIP2010}, they have become widely used in applications in remote sensing~\cite{Stein2002}, medical diagnosis~\cite{Medicalhyper2014}, face recognition~\cite{facehyperTIP2015,HyperfaceTPAMI2003}, and quality control~\cite{GENDRIN2008}.
However, real HSIs often suffer from different kinds of degradations, \textit{i.e.,} noise~\cite{chang2017hyper,Dong2015ICCV,CVPR2014Meng,xie2017kronecker},
undersampling~\cite{martin2014hyca,wang2016adaptive,zhang2018cluster} or missing data~\cite{he2019total,ng2017adaptive,xie2018tensor},
because of imaging conditions in practice, weather conditions, or data transmission procedures. These kinds of degradation substantially influence the subsequent processing,
and as a result, HSI restoration is a fundamental initial step for quality improvement and subsequent exploitation~\cite{zhang2018cluster,He2014TGRS,changyiTIP2015}.

Restoration from a noisy, undersampled, or incomplete HSI is an ill-posed inverse problem,
and the prior, which is also called a regularizer, is necessary to constrain the solution space~\cite{xie2017kronecker,Zhang_2017_CVPR,zhang2018cluster}.
HSIs have redundant information that can be regularized as different priors for the solution space in HSI restoration.
The relevant popular priors can be categorized as spatial correlation~\cite{yuan2016generalized,chang2017hyper} and spectral correlation~\cite{He2014TGRS}.
Traditional spatial-correlation regularized methods, such as total variation~\cite{yuan2016generalized}, can be regarded as an extension of color image restoration.
However, these methods cannot characterize the main features of HSI and fails to achieve state-of-the-art restoration results~\cite{changyiTIP2015,yuan2016generalized}.
In this paper, we focus on one specific kind of spatial correlation, named non-local similarity.
%However, spatial-related strategies are always adopted as s supporting technique to further improve the restoration accuracy of spectral correlation and non-local similarity related methods.
Methods based on non-local similarity have achieved state-of-the-art performance in HSI restoration tasks, such as denoising \cite{chang2017hyper,Dong2015ICCV,CVPR2014Meng,xie2017kronecker,baiJSTAR2018},
compressed HSI reconstruction~\cite{wang2016adaptive,zhang2018cluster,DongCS2014,lishutaoCVPR2017,DianTCYB}, and
inpainting~\cite{ng2017adaptive,xie2018tensor}.
For different restoration tasks, a non-local strategy processes the HSIs via group matching the \textit{full band patches}
(FBPs, which are stacked by patches at the same HSI location over all bands) and a low-rank approximation of each non-local FBP group (NLFBPG).
This kind of non-local processing strategy faces a crucial problem. For HSIs, a higher number of spectral dimensions means a higher discriminant ability~\cite{Bioucas2012jstars},
thus more spectra are desired for subsequent application.
However, as the number of spectra increases,
the size of each NLFBPG also increases,
leading to substantially more computations for the subsequent low-rank matrix/tensor approximations.

The spectral correlation of HSIs, modeled as a spectral low-rank approximation, has also been widely utilized for HSI denoising \cite{He2014TGRS,xie2016hyperspectral}, compressed HSI reconstruction \cite{khan2015joint,fan2015exploiting,waters2011sparcs} and inpainting \cite{ng2017adaptive,he2019total,xu2013parallel,zhuang2018fast,yao2019efficient}. However, because of the lack of spatial regularization, spectral low-rank regularization alone cannot restore the HSI efficiently. Therefore, spatial-based methods are embedded into the spectral low-rank regularization to simultaneously restore the HSI \cite{HE2016TGRS,lu2013graph,fu2017adaptive,golbabaee2012joint,peng2018enhanced,wang2017compressive}. Another promising improvement is to project the original degraded HSI onto a low-dimensional spectral subspace
and restore the projected HSI via spatial-based methods~\cite{rasti2014hyperspectral,zhuang2018fast}.
Unfortunately,
these two-stage methods are heavily influenced by the quality of projection and the efficiency of spatial restoration.
All of them fail to capture a clean projection matrix,
which keeps the low quality of the restored HSI.

\begin{figure*}[t]
	\centering
	\includegraphics[width= 0.9 \linewidth]{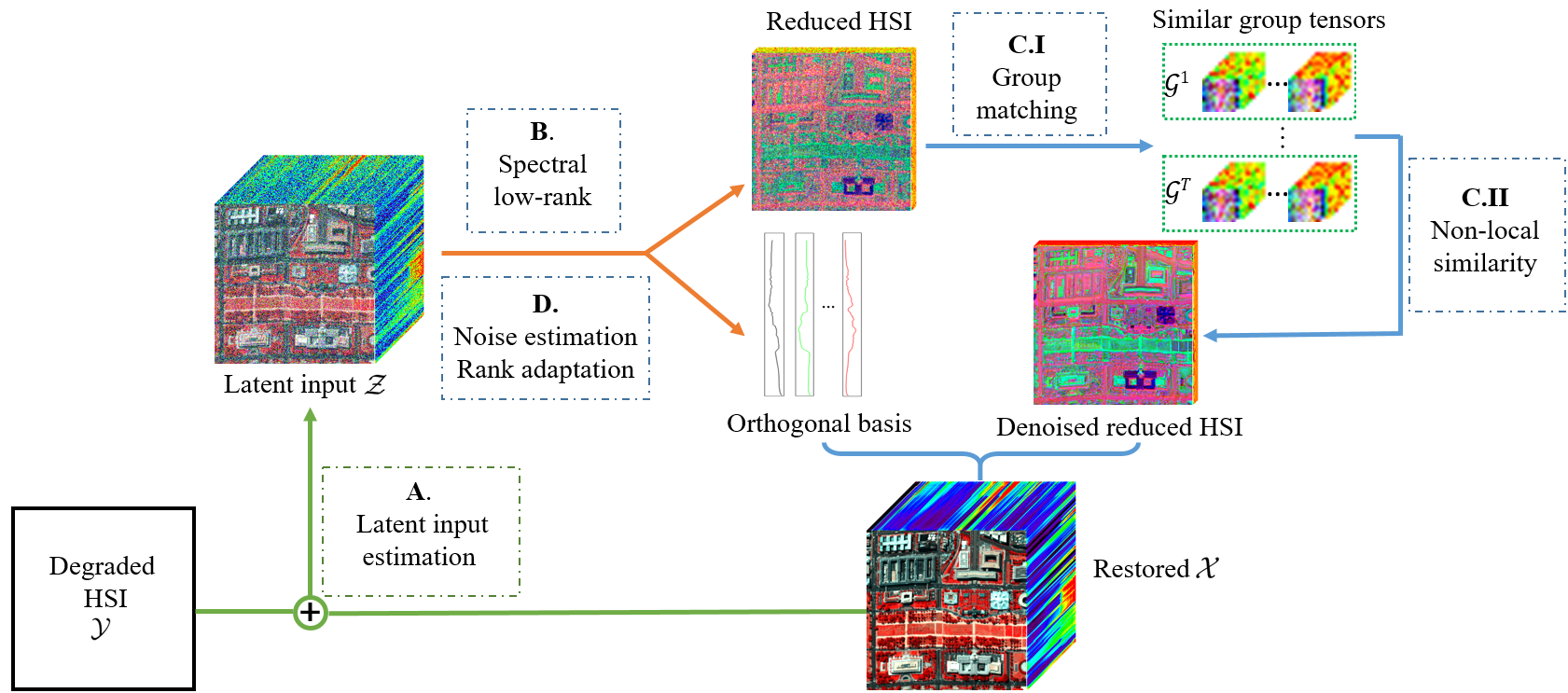}
	\caption{Flowchart of the proposed method (Algorithm~\ref{alg:NSLR_base}).
		It includes: A. latent input HSI estimation, B. Spectral orthogonal basis optimization and C. non-local similarity estimation. C consists of two steps including group matching and non-local low-rank approximation. D. Noise estimation and rank adaptation.}
	%%\vspace{-10px}
	\label{fig:flow_m1}
\end{figure*}

This paper proposes a unified paradigm to deal with various HSI restoration tasks, namely, denoising, compressed HSI reconstruction, and inpainting. To alleviate the aforementioned problems, the proposed paradigm integrates the spatial non-local similarity with the global spectral low-rank property. We start from the viewpoint that the HSI should lie in a low-dimensional spectral subspace~\cite{dian2019hyperspectral,dian2019learning}, which has been widely accepted for hyperspectral imaging~\cite{FuCVPR2016}, denoising~\cite{MengTIP2016}, and compressed sensing~\cite{MengTIP2016,weiwei2015CVPR} tasks. Based on this fact, the all NLFBPGs should also lie in a common low-dimensional spectral subspace.
%\footnote{+++ cannot understand... better to break this sentence into two shorter sentences.{\color{red} changed}}
Thus, the proposed paradigm first learns a global spectral low-rank orthogonal basis and subsequently exploits the spatial non-local similarity of the projected HSI on this basis. \textit{In our paradigm, the computational cost of non-local processing remains almost the same even with more spectral bands, and the global spectral low-rank property is also enhanced.}
The main novelties and contributions are summarized as follows:
\footnote{Code is avaliable at: \url{https://github.com/quanmingyao/NGmeet}}

\begin{itemize}[leftmargin = 10pt]
\item We propose a new unified paradigm for HSI restoration, that can \textit{jointly learn} and \textit{iteratively update} the orthogonal basis matrix and reduced image
%\footnote{{\color{red}+++ this figure should not focus on denoising now.}}
(Fig.~\ref{fig:flow_m1}).
The proposed paradigm can guarantee the performance superiority via state-of-the-art non-local denoising methods, meanwhile get rid of the considerable computation burden caused by high spectral dimension non-local processing.

\item An efficient alternating minimization algorithm with convergence analysis is developed. To enhance the real applications, we also design a linear strategy to adaptively predict the dimension of the orthogonal basis.
%The proposed new model for HSI restoration is challenging to optimize
%because it involves both complex constraint
%(from the orthogonal basis optimization) and regularization
%(from the spatial restoration).
%We further propose an efficient and iterative algorithm for optimization that is inspired by alternating minimization approach.
\item Finally,
the proposed restoration method is evaluated on simulated and real data. It is compared with other state-of-the-art methods on three important HSI
restoration tasks,
i.e., denoising, compressed HSI reconstruction, and inpainting.
%and real data experiments demonstrate the advantage of the proposed method.
\end{itemize}

This paper is an extension of a previous conference paper \cite{he2018non}.
Compared to~\cite{he2018non}, we extend the work as follows:
(i) We introduce a new objective (Section~\ref{sec:framework}),
which extends our model from HSI denoising to a more general HSI restoration model, that includes denoising, compressed HSI reconstruction, and inpainting (Section~\ref{sec:application}).
(ii) The optimization in our conference paper is specific to denoising, as iterative regularization is used.
Here, we propose an efficient alternating minimization algorithm with rank adaptation to solve the objective (Section~\ref{sec:opt}).
While the proposed algorithm is still based on alternating minimization, its subproblems are different, and it depends on a restoration model.
(iii) We conduct additional experiments to evaluate HSI restoration, and our unified model achieves the best results not only for HSI denoising,
but also for the compressed HSI reconstruction (Section~\ref{sec:exp:comp}) and inpainting (Section~\ref{sec:imp}) tasks.

\subsubsection*{Notations}
We follow the tensor notation in \cite{Kolda2009},
the tensor and matrix are represented as Euler script letters, \textit{i.e.} $\mathcal{X}$ and boldface capital letter, \textit{i.e.} $\mathbf{A}$, respectively.
%Tensor unfolding is the process to reorder the elements of a tensor into a matrix.
For a $N$-order tensor $\mathcal{X}\in\mathbb{R}^{I_1\times I_2\times\cdots \times I_N}$,
the mode-$n$ unfolding operator is denoted as $\mathbf{X}_{(n)}\in\mathbb{R}^{I_n \times  {I_1 \cdots I_{n-1} I_{n+1} \cdots I_N}}$.
We have $\text{fold}_n(\mathbf{X}_{(n)})=\mathcal{X}$, in which $\text{fold}_n$ is the inverse operator of unfolding operator. The Frobenius norm of $\mathcal{X}$ is defined by $\left \| \mathcal{X} \right \|_F= (\sum_{i_1}\sum_{i_2}\cdots\sum_{i_N}x_{i_1 i_2\ldots i_N}^2)^{0.5}$.
The mode-$n$ product of a tensor $\mathcal{X}\in\mathbb{R}^{I_1\times I_2\times\cdots \times I_N}$ and a matrix
$\mathbf{A}\in\mathbb{R}^{J_n\times I_n}$ is defined as $\mathcal{Y} = \mathcal{X} \times_n
\mathbf{A}$, where $\mathcal{Y}\in\mathbb{R}^{I_1\times I_2\times\cdots \times J_n}$ and $\mathcal{X}\times_n \mathbf{A} = \text{fold}_n(\mathbf{A} \mathbf{X}_{(n)})$.

\section{Related work}
\label{sec:rel}

Because restoration is an ill-posed problem,
proper regularization from HSI prior knowledge is necessary~\cite{fu2017adaptive}.
The mainstream regularizer for HSI restoration can be grouped into two categories:
non-local similarity regularizers and low-rank regularizers.
%\footnote{{\color{red}+++ I cannot see this point here.
%	Currently,
%	it is still in the same form of CVPR.
%	It is better to make more difference now.}}
We review these methods for HSI restoration in this section.

\subsection{Non-local similarity based methods}
\label{sec:rel:local}

Non-local similarity regularization has been widely utilized in HSI denoising~\cite{chang2017hyper},
compressed HSI reconstruction~\cite{xue2019nonlocal,Yuan_PAMI_2019} and inpainting~\cite{xie2018tensor}.
Specifically,
\cite{CVPR2014Meng} was the first to introduce non-local low-rank modeling for HSI denoising.
this non-local denoising architecture has two stage: FBPs grouping and low-rank tensor approximation.
On the basis of this architecture, different kinds of low-rank modeling for NLFBPGs were proposed to simultaneously exploit the spatial non-local similarity and spectral low-rank property,
including Tucker decomposition~\cite{CVPR2014Meng}, sparsity regularized Tucker decomposition~\cite{xie2017kronecker},
Laplacian scale mixture low-rank modeling~\cite{Dong2015ICCV}, weighted nuclear norm minimization~\cite{Yuan_PAMI_2019}, dimension-discriminative low-rank tensor recovery~\cite{zhang2019computational}, weighted low-rank tensor recovery \cite{chang2017weighted}, and tensor ring decomposition~\cite{Yong2020TIP}.
By estimating the latent low quality HSI from the degraded observations,
the compressed HSI reconstruction and inpainting tasks are converted to the non-local denoising of a latent low quality HSI,
resulting in state-of-the-art reconstruction~\cite{fu2017adaptive,zhang2018cluster,wang2016adaptive,xue2019nonlocal} and inpainting~\cite{DongCS2014,ji2018nonlocal,xie2018tensor} results.
However, as the number of spectra increases, the computational burden also increases substantially, impeding the application of non-local methods to real high-dimensional HSIs.

The above methods adopt low-rank regularization to explore the prior of local spatial correlation. However, Chang et al.~\cite{chang2017hyper} claimed that the spectral and local low-rank property of NLFBPGs is weak. Hence, they simply adopt a non-local low-rank prior to produce state-of-the-art HSI denoising results, while substantially reduce the computational burden.
Thus, previous methods sub-optimally utilize spectral and local spatial correlation.
How to balance the contributions of spectral correlation and non-local similarity remains a problem.

\subsection{Global low-rank based methods}
\label{sec:rel:global}

Global spectral low-rank correlation is another powerful prior for HSI restoration. As described in~\cite{BioucasTGRS2008},
the HSI has approximate spectral low-rank property, and the approximate dimension of the spectral subspace can be far less than that of the original image.
Thus, by vectorizing each band of the HSI and reshaping the original 3D HSI into a 2D matrix, various low-rank approximations such as principal component analysis (PCA)~\cite{khan2015joint,arce2013compressive,fowler2009compressive},
robust PCA~\cite{He2014TGRS,xie2016hyperspectral,waters2011sparcs}, and
low-rank matrix factorization~\cite{MengTIP2016} have been directly adopted to the restoration of HSI. However, the reshaping processing of the spatial dimensions destroys the spatial correlation. Low-rank tensor approximation is one way to additionally exploit the spatial low-rank correlation~\cite{renard2008denoising,xu2013parallel,ng2017adaptive,gong2020low,chen2019hyperspectral}. However, as pointed out in the introduction, simple low-rank regularization for spatial correlation is inappropriate, resulting in the failure of such methods to achieve the optimal restoration performance. Typically, there are two strategies to combine spatial correlation and the spectral low-rank property.
In the first approach, many conventional spatial regularizers such as total variation~\cite{cao2016total,golbabaee2012joint,HE2016TGRS,ji2016tensor} are combined with low-rank matrix/tensor approximation to simultaneously exploit the spatial and spectral properties. Alternatively, a two-stage method is used to combine the spatial regularizer and spectral low-rank property together. This is done by first mapping the original HSI into a low-dimensional spectral subspace, and then restoring the mapped image via existing spatial restoration method or HOSVD~\cite{zhuang2017hyperspectral,cao2019hyperspectral}.
These two-stage methods provide a new insight to restore the HSI in the transferred spectral space that is very fast.
However,
these methods do not refine the subspace
and thus fail to combine the best of both worlds, \textit{i.e.,} non-local similarity and the global spectral low-rank property. Moreover, the extracted subspace is still of low quality.

\section{Proposed NGmeet Paradigm}

%\footnote{ {\color{red} +++ if we claim it as a regularizer,
%	then we'd better to discuss related works from regularizers' perspective.}}
In this section, we propose a unified HSI restoration paradigm to integrate
the spatial non-local similarity prior
and global spectral low-rank property prior. We first solve a fidelity term-related problem to learn the latent input HSI by enhancing the consistency of the recovered output image with the observed data,
and then learn a low-dimensional orthogonal basis and the related reduced image from the latent input HSI to explore the global spectral low-rank prior.
Subsequently, the reduced image is updated by the non-local similarity prior. Finally, the optimization is iterated to refine the orthogonal basis and the reduced image, alternately.
An overview of the proposed restoration paradigm is shown in Fig.~\ref{fig:flow_m1}.

\subsection{Unified restoration paradigm}
\label{sec:framework}

Summarized from~\cite{xie2017kronecker,zhang2018cluster,xie2018tensor,Yuan_PAMI_2019},
we assume that a clean HSI $\mathcal{X} \in \mathbb{R}^{M \times N \times B}$
is corrupted by linear degradation operator $h$ and
additive Gaussian noise $\mathcal{N}$ (with zero mean and variance $\sigma_0^2$).
Then the degraded HSI $\mathcal{Y}$ is generated by
\begin{equation}
\mathcal{Y} = h (\mathcal{X}) + \mathcal{N}.
\label{eq:basic}
\end{equation}
By specifying different $h$,
\eqref{eq:basic} can be adapted for various HSI restoration tasks.
In this study, we choose three classical HSI tasks, \textit{i.e.,}
denoising~\cite{xie2017kronecker}
where $h$ is an identity operator (Section~\ref{sec:denoise}),
reconstruction~\cite{zhang2018cluster,Yuan_PAMI_2019}
where $h$ is a compressed measurement operator (Section~\ref{sec:compress}),
and inpainting~\cite{he2019total,xie2018tensor,yao2019efficient}
where $h$ is a sampling operator (Section~\ref{sec:impaint}).
Image restoration recovers a desired unknown image from corrupted observations given by \eqref{eq:basic}.
Usually,
$\mathcal{Y}$ is a sampled vector in the compressed HSI reconstruction problem and a tensor in HSI denoising and inpainting problems.
Because the corruption process is irreversible,
image restoration is usually an ill-posed problem \cite{xie2017kronecker,zhang2017learning}.
Thus,
it is necessary to utilize prior information or a regularizer for image restoration
\cite{xie2017kronecker,zhang2018cluster}.

In this paper,
%\footnote{+++ [to quanming] check related work later}
motivated by the problems with existing works discussed in Section~\ref{sec:rel},
we propose a unified HSI restoration paradigm to simultaneously capture the non-local similarity and global spectral low-rank property priors.
%\footnote{{\color{red}+++ contributions in the conference version also counts.}}
%First, to capture the spectral low-rank property in Section~\ref{sec:rel:global},
%we propose a spatial-spectral regularizer for the regularization of HSI restoration \eqref{eq:globopt}.
\begin{itemize}[leftmargin = 10pt]
\item First, to capture the spectral low-rank property (Section~\ref{sec:rel:global}),
we are motivated to use a low-rank representation on the clean image,
\textit{i.e.},
\begin{align}
\label{eq:temp1}
\mathcal{X} = \mathcal{M} \times_3 \mathbf{A},
\end{align}
where $K \ll B$, $\mathbf{A}\in \mathbb{R}^{B \times K}$ is an orthogonal basis matrix capturing the common subspace of different spectra,
and $\mathcal{M}\in \mathbb{R}^{M \times N \times K}$ is the reduced image.
%\footnote{{\color{blue}Revision+++ I feel the meaning is still not exact here.
%		1). the regularizer imposes some prior on a group of image patches;
%		2). the group is pre-defined, not learned by the regularizer;
%		3). low-rank is not a must, TV and sparse coding is also OK (mathematically).}}
%\footnote{{\color{blue}+++ cont. above footnote;
%		we can make ``Non-local similarity'' more specific here or in previous Section~2.1.}}
%This formulation \eqref{eq:NLdenoising} appears in many denoising models,
%\textit{e.g.} WNNM~\cite{gu2014weighted},
%TV \cite{HE2016TGRS}, wavelets \cite{rasti2014hyperspectral} and CNN \cite{chang2018hsi}.
%Specifically,
%to solve this regularizer,
%we need to first group similar patches,
%then denoise the tensors of each patch group,
%and then assemble the final estimated $\mathcal{M}_i$.

%to utilize the spatial low-rank property (Section~\ref{sec:rel:local}),
%we add a non-local low-rank regularizer $\| \cdot \|_{\text{NL}}$ on the reduced image $\mathcal{M}$.

\item
Second, we add a spatial regularizer to denoise the reduced image $\mathcal{M}$.
In this paper, we follow the state-of-the-art architecture as in \cite{chang2017hyper,Yuan_PAMI_2019} and adopt a non-local regularizer
\begin{align}
\label{eq:NL_define}
\| \mathcal{M} \|_{\text{NL}}= \sum\nolimits_j
%\lambda_j \;
r( \left[ \mathcal{M} \right]_{\mathcal{G}^j} ),
\end{align}
where $\mathcal{G}^j$ represents the pre-defined non-local group,
$\left[ \mathcal{M} \right]_{\mathcal{G}^j}$ extracts corresponding patches indicated by $\mathcal{G}^j$ from the image $\mathcal{M}$ to form the $j$-th exemplar patch group,
%$\lambda_j$ is the hyper-parameter for the $j$th patch group,
and $r$ is a regularizer imposed on all patches groups.
Typically, TV \cite{HE2016TGRS},
wavelets \cite{rasti2014hyperspectral} and CNN \cite{chang2018hsi,DianTNNLS} can be adopted as the spatial regularizer $r$ to reduce the noise of $\mathcal{M}$.

%$\mathcal{M}^j$ standing for the $j$-th exemplar patch of variable $\mathcal{M}$, and $\Vert \cdot \Vert_{reg}$ standing for the regularizer of each non-local groups.
%In this paper, we adopt nuclear norm $\| \cdot \|_{*}$, i.e., the sum of singular values.
\end{itemize}
Putting \eqref{eq:temp1} and \eqref{eq:NL_define} into \eqref{eq:basic},
the proposed non-local meets global (NGmeet) restoration paradigm is represented as
%\footnote{+++ need to explain why we do not write it as
%	$\Vert \mathcal{X} - \mathcal{M} \times_3 \mathbf{A} \Vert_F^2+ \mu{\Vert \mathcal{M} \Vert_{\text{NL}}}$,
%	note that in (9), we actually implement it in this way.}
\begin{align}
\min_{\mathbf{A}, \mathcal{M}}
\frac{1}{2}
\left\| \mathcal{Y} \! - \! h(\mathcal{M} \! \times_3 \! \mathbf{A} ) \right\|_F^2
\! + \! \lambda \left\| \mathcal{M} \right\|_{\text{NL}}
\;\text{s.t.}\;
\mathbf{A}^{\top} \mathbf{A} = \mathbf{I},
\label{eq:overall}
\end{align}
where $\lambda$ is the trade-off parameter between the recovered output image and observed data.
In addition, basis matrix $\mathbf{A}$ must be orthogonal.
%and the output HSI is recovered by $\mathcal{M} \times_3 \mathbf{A}$.

%\begin{align*}
%\left\| \mathcal{M} \right\|_{\text{NL}}
%= \sum_{g = 1}^M \| \mathcal{P} \|
%\end{align*}

\begin{remark}
The orthogonal constraint $\mathbf{A}^{\top} \mathbf{A} = \mathbf{I}$ is very important here.
%\footnote{+++ we need to add experiments to show this point}
First,
it encourages the representations held
in $\mathbf{A}$ to be more distinct.
This helps to keep noise out of $\mathbf{A}$.
% and further allows a closed-form solution for computing $\mathbf{A}$
%\footnote{+++ need changes.}
%(Section~\ref{sec:opt:spectral}).
It preserves the distribution of noise,
which allows us to estimate the remaining noise levels in the reduced image and
employ a state-of-the-art Gaussian based non-local method for spatial denoising \eqref{eq:NLdenoising}.
\end{remark}

\subsection{Efficient optimization by alternative minimization}
\label{sec:opt}

The objective \eqref{eq:overall} needs to handle the orthogonal constraint on $\mathbf{A}$ and
complex regularization on $\mathcal{M}$.
Inspired by the success of half-quadratic splitting in image processing \cite{nocedal2006numerical,zhang2017learning},
we introduce an auxiliary variable $\mathcal{Z}$ and propose to relax \eqref{eq:overall} as follows:
\begin{align}
\label{eq:overall_L}
\left\lbrace \mathbf{A}^*, \mathcal{M}^*, \mathcal{Z}^* \right\rbrace
= & \underset{\mathbf{A}, \mathcal{M}, \mathcal{Z}}{\arg\min} F(\mathbf{A}, \mathcal{M}, \mathcal{Z})
\end{align}
where $F$ is defined as
\begin{align*}
F(\mathbf{A}, \mathcal{M}, \mathcal{Z})
& = \frac{1}{2}
\left\| \mathcal{Y} \! - \! h( \mathcal{Z} ) \right\|_F^2
+ \lambda \left\| \mathcal{M} \right\|_{\text{NL}}
\\
& + \frac{\mu}{2} \left\| \mathcal{Z}  - \mathcal{M} \times_3 \mathbf{A} \right\|_F^2,
\;\text{s.t.}\;
\mathbf{A}^{\top} \mathbf{A} = \mathbf{I},
\end{align*}
and the last quadratic term encourages
$\mathcal{Z}$ to be closed to $\mathcal{M} \times_3 \mathbf{A}$,
and $\mu$ is a hyper-parameter to be tuned.

%We adopt $\mathcal{Z}\times_3 \mathbf{A}^{\top} - \mathcal{M}$ instead of $\mathcal{Z} - \mathcal{M}\times_3 \mathbf{A}$ to facilitate the optimization of the reduced image $\mathcal{M}$ in \eqref{eq:NLdenoising} {\color{red}and linearly increase the dimension of orthogonal basis matrix $\mathbf{A}$ via \eqref{delta}}.

%\subsubsection{Alternative Minimization}
Based on the reformulation in \eqref{eq:overall_L},
we are motivated to use alternating minimization
\cite{nocedal2006numerical,wang2008new}
for optimization
(Algorithm~\ref{alg:NSLR_base}). Let $\mathcal{Z}_i$ and $\mathcal{X}_i$ stand for the latent input image and the recovered output image of the $i$-th iteration, respectively.
The aim of Algorithm~\ref{alg:NSLR_base} is to update the latent input image (step-3) to enhance the consistency of the recovered output image with respect to observed data $\mathcal{Y}$. To further improve the quality of latent input image $\mathcal{Z}_i$, we find an optimization for $\mathbf{A}$ (step~4) to exploit the spectral low-rank property,
and use a state-of-the-art non-local denoising method to compute $\| \cdot \|_{\text{NL}}$ (step~5).
% The purpose here is to utilize the spatial non-local similarity.

%\begin{algorithm}[ht]
%\caption{Baseline Non-local Meets Global (NGmeet).}
%\label{alg:NSLR_base}
%\begin{algorithmic}[1]
%	\REQUIRE Observed image $\mathcal{Y}$, noise variance $\sigma_0^2$ (for denoising), dimension $K$, initialized $\mathcal{X}_0$;
%	\STATE Initializing $\mathcal{X}_0$, the pre-defined non-local groups {\{$\mathcal{G}^j\}$};
%	\FOR{$i = 1, 2, 3, \cdots iter$}
%
%    \STATE{A). \textit{Latent input image estimation via \eqref{eq:sub1}};}
%    \\
%    {Estimate latent input image $\mathcal{Z}_i$;}
%	\STATE {B). \textit{Spectral orthogonal optimization optimization via} \eqref{eq:sub:a_new}}:
%	\\
%	 {Estimate orthogonal basis matrix $\mathbf{A}_i$;}
%	\STATE {C). \textit{Non-local denoising on $\bar{\mathcal{M}}_i$} via \eqref{eq:NLdenoising}:}
%	\\
%      {Denoise groups $\{ \left[ \bar{\mathcal{M}}_i \right]_{\mathcal{G}^j} \}$ via Low-rank approximation, obtain $\{\mathcal{M}^j\}$ and reconstructed image $\mathcal{M}_i$;}
%	\ENDFOR
%	
%	\RETURN  Recovered image $\mathcal{X}_i= \mathcal{M}_{i}\!\times_3\! \mathbf{A}_{i}$.
%\end{algorithmic}
%\end{algorithm}

\begin{algorithm}[ht]
	\caption{Non-local Meets Global (NGmeet).}
	\label{alg:NSLR_base}
	\begin{algorithmic}[1]
		\REQUIRE Observed image $\mathcal{Y}$, noise variance $\sigma_0^2$
		\STATE Initializing $\mathcal{X}_0$, estimating $K$ using HySime~\cite{BioucasTGRS2008};
		\FOR{$i = 1, 2, 3, \cdots iter$}
		
		\STATE{A). \textit{Latent input image estimation} via \eqref{eq:sub1};}
		\\
		{Estimate latent input image $\mathcal{Z}_i$;}
		\STATE {B). \textit{Spectral orthogonal optimization optimization via \eqref{eq:diclearning}}}:
		\\
		{Estimate orthogonal basis matrix $\mathbf{A}_i$ and reduced image $\bar{\mathcal{M}}_i$ via SVD on $\mathcal{Y}_i$;}
		\STATE {C). \textit{Non-local denoising on $\bar{\mathcal{M}}_i$} via \eqref{eq:NLdenoising}:}
		\\
		{-C.I) Obtain the set of non-local groups $\left\{\mathcal{G}_i^j\right\}$
			for $\bar{\mathcal{M}}_i$ via $k$-NN search for each reference patch in \eqref{eq:proxav1};} \\
		{-C.II) Denoise $\{ \left[ \bar{\mathcal{M}}_i \right]_{\mathcal{G}_i^j} \}$ via Low-rank approximation and obtain $\{\mathcal{M}^j\}$ with $\sigma_i^2$ in \eqref{eq:proxavg2};} \\
		{-C.III) Reconstruct the cubes $\{ \mathcal{M}^j \}$ to image $\mathcal{M}_i$,
			and obtain the denoised HSI $\mathcal{X}_i=\mathcal{M}_i\times_3\mathbf{A}_i$ in \eqref{eq:proxav3};}
		
		\STATE {-D.I). \textit{Noise re-estimation:}}  { via \eqref{eq:noise-estimation};}
		\\
		{-D.II). \textit{Rank adaptation:}}  {${K=min(K+\delta \times i,B)}$;}
		\ENDFOR
		
		\RETURN  Recovered image $\mathcal{X}_i= \mathcal{M}_{i}\!\times_3\! \mathbf{A}_{i}$.
	\end{algorithmic}
\end{algorithm}

 We provide the theoretical convergence analysis for Algorithm~\ref{alg:NSLR_base} in Section \ref{the:conv}, by fixing the ideal cases of noise variance $\sigma_i^2$, dimension $K$, and non-local groups $\left\{\mathcal{G}^j\right\}$ for each iteration. However, in the real applications, the estimation of dimension $K$ and the ideal non-local similar groups are the key problems. We introduce some empirical enhancement including non-local similar groups re-matching (Step~5-C.I), noise re-estimation (Step~6-D.I), and dimension $K$ re-estimation (Step~6-D.II) for each iteration to make sure that  NGmeet is more efficient for the real HSIs restoration. Next, we present the details to each step in Algorithm~\ref{alg:NSLR_base}.

\subsubsection*{Step~3: Latent input image estimation}
We first fix variables $\mathcal{M}_{i-1}$ and $\mathbf{A}_{i-1}$ and update latent variable $\mathcal{Z}_i$.
The optimization is formulated as
\begin{align}
\mathcal{Z}_i
\! = \!
\arg\min_{\mathcal{Z}}\!
\frac{1}{2}\Vert \mathcal{Y} \! - \! h(\mathcal{Z}) \Vert_F^2
\! + \! \frac{\mu}{2} \left\| \mathcal{Z} - \! \mathcal{M}_{i - 1}  \!\times_3\! \mathbf{A}_{i - 1} \! \right\|_F^2.\!\!
\label{eq:sub1}
\end{align}
Specifically, the optimization of \eqref{eq:sub1} is related to a quadratic regularized least-square problem,
which has different fast solutions for different degradation problems.
We discuss the solution for different HSI restoration tasks in Section \ref{sec:application}.

\subsubsection*{Step~4: Spectral orthogonal basis optimization via $\mathbf{A}$}
In this stage,
we identify the orthogonal basis matrix $\mathbf{A}$ with
the given $\mathcal{M}_{i-1}$ and $\mathcal{Z}_i$ from \eqref{eq:overall_L},
which leads to
\begin{align}
\mathbf{A}_i
& =\underset{\mathbf{A}^{\top} \mathbf{A} = \mathbf{I}}{\arg\min}
\frac{1}{2} \Vert \mathcal{Z}_{i} - \mathcal{M}_{i - 1} \times_3 \mathbf{A}\Vert_F^2.
%\notag
%\\
%& = \underset{\mathbf{A}^{\top} \mathbf{A} = \mathbf{I}}{\arg\max}
%\left\langle \mathbf{A},(\mathbf{Z}_{i})_{(3)}(\mathbf{M}_{i - 1})_{(3)}^{\top} \right\rangle.
\label{eq:sub:a_new}
\end{align}
%%%%%%%
The key challenge to optimizing $\mathbf{A}$ is that, as we increase the dimension $K$ of $\mathbf{A}_{i+1}$ in the next \textit{i+}1 iteration, we will lose the closed-form solution of optimizing $\mathbf{A}$ via \eqref{eq:sub:a_new}. Thus,
we propose to relax \eqref{eq:sub:a_new} as
\begin{equation}
\{ \bar{\mathcal{M}}_i, \mathbf{A}_i \}
= \underset{\mathcal{M},\mathbf{A}^{\top} \mathbf{A}=\mathbf{I}}{ \arg \min }
\frac{1}{2}
\| \mathcal{Z}_i - \mathcal{M}\times_3\mathbf{A} \|_F^2.
\label{eq:diclearning}
\end{equation}
%\footnote{$\surd$ quanming:{\color{blue} As suggest by the reviewer, I've removed the proposition.}}
The closed-form solution to \eqref{eq:diclearning}
can be obtained by a singular value decomposition (SVD) on the folding matrix of $(\mathcal{Z}_i)_{(3)}$, which can be efficiently computed. The optimization of $\mathbf{A}$ via \eqref{eq:diclearning} is simply related to the latent input image $\mathcal{Z}_i$ and suitable for any $K$.

\begin{remark}
	We also provide a more elegant solution to \eqref{eq:sub:a_new}
	\begin{align}
	\mathbf{A}_i
	& =\underset{\mathbf{A}^{\top} \mathbf{A} = \mathbf{I}}{\arg\min}
	\frac{1}{2} \Vert \mathcal{Z}_{i} - \mathcal{M}_{i - 1} \times_3 \mathbf{A}\Vert_F^2
	\notag
	\\
	& = \underset{\mathbf{A}^{\top} \mathbf{A} = \mathbf{I}}{\arg\max}
	\left\langle \mathbf{A},(\mathbf{Z}_{i})_{(3)}(\mathbf{M}_{i - 1})_{(3)}^{\top} \right\rangle.
	\label{eq:sub:a_new_r3}
	\end{align}
	According to \cite{xie2017kronecker}, the optimization of $\mathbf{A}$ via \eqref{eq:sub:a_new_r3} has the closed-form solution of $\mathbf{A}_i=\mathbf{B}\mathbf{C}^{\top}$, where $(\mathbf{Z}_{i})_{(3)}(\mathbf{M}_{i - 1})_{(3)}^{\top} = \mathbf{B}\mathbf{D}\mathbf{C}^{\top}$ is the SVD. However, the optimization \eqref{eq:sub:a_new_r3} is only suitable for the first iteration. In the next iteration, the size of $\mathbf{A}_{i+1}$ remains the same, as the limitation of the size of $\mathcal{M}_{i}$. Therein, an empirically enhancement is introduced to augment the spectral size of $\mathcal{M}_{i}$ via $[\mathcal{M}_{i}, \mathcal{N}_{i}]$, where $\mathcal{N}_{i} \in \mathbb{R}^{M \times N \times \delta}$, and $(\mathbf{N}_{i})_{(3)}$ can be achieved by the first $\delta$ right singular value vectors of $(\mathcal{Z}_{i+1} - \mathcal{M}_{i} \times_3 \mathbf{A_{i}})_{(3)}$. The meaning of $\delta$ will be introduced in \eqref{delta}.
	Furthermore, the optimization of $\mathbf{A}$ via \eqref{eq:sub:a_new_r3} and \eqref{eq:diclearning} yields nearly identical HSI restoration results as analyzed in Section~\ref{Empirical:analysis}.
\end{remark}

\subsubsection*{Step~5: Non-local denoising on $\bar{\mathcal{M}}_i$}

Here, we fix $\mathcal{A}_{i}$ and $\mathcal{Z}_i$,
and identify the reduced image $\mathcal{M}$ from \eqref{eq:overall_L}, which is given by
\begin{align}
\label{eq:NLdenoising}
\underset{\mathcal{M}}{\arg\min}
\frac{1}{2} \left\| \bar{\mathcal{M}}_i - \mathcal{M} \right\|_F^2 +
\lambda_0 \left\| \mathcal{M} \right\|_{\text{NL}}.
\end{align}
where $\bar{\mathcal{M}}_i = \mathcal{Z}_{i}\times_3 \mathbf{A}_{i}^{\top}$
is from Lemma \ref{lemma1} and step 4, and $\lambda_0 = \nicefrac{\lambda}{\mu}$. % and $\Vert \cdot \Vert_{\text{NL}}$ is a non-local denoising regularizer.

\begin{lemma}
	\label{lemma1}
	$\arg\min_{\mathcal{M}}
	\frac{1}{2}\left\| \mathcal{Z} \! - \! \mathcal{M}\times_3 \mathbf{A} \right\|_F^2 \! + \!
	\frac{\lambda}{\mu}\left\| \mathcal{M} \right\|_{\text{NL}}
	\!= \! \arg\min_{\mathcal{M}}\frac{1}{2}\left\| \mathcal{Z}\times_3 \mathbf{A}^{\top} \! - \! \mathcal{M}\right\|_F^2 \! + \!
	\frac{\lambda}{\mu}\left\| \mathcal{M} \right\|_{\text{NL}}$ when $\mathbf{A}^{\top} \!\! \mathbf{A} \! = \! \mathbf{I}$.
\end{lemma}

With \eqref{eq:NLdenoising}, we transfer the high dimensional original HSI denoising to the low-dimensional reduced image $\bar{\mathcal{M}}_i$ denoising.
However,
since patches are overlapped with each other,
there is no simple closed-form solution for \eqref{eq:NLdenoising}.
Following other works on non-local image denoising \cite{rudin1992nonlinear,wang2008new,mairal2009non,dong2013nonlocal,gu2014weighted,zhuang2018fast,xie2018tensor},
we also approximate the optimal solution of \eqref{eq:NLdenoising} via following three steps
\begin{align}
\mathcal{A}_i^j
& = \left[ \bar{\mathcal{M}}_i \right]_{\mathcal{G}_i^j} \text{for all groups},
\label{eq:proxav1}
\\
\mathcal{M}^j
& = \Px{\lambda_0 r}{\mathcal{A}_i^j}
%\underset{\mathcal{A}}{\arg\min}
%\frac{1}{2} \left\| \mathcal{A}^j \! - \! \mathcal{A} \right\|_F^2 \! + \lambda r(\mathcal{A}),
\;\text{for all groups},
\label{eq:proxavg2}
\\
\mathcal{M}^*
& = \text{aggrate all}\; \mathcal{M}^j,
\label{eq:proxav3}
\end{align}
where
	\eqref{eq:proxav1} represents the searching for non-local groups $\{\mathcal{G}_i^j\}$ as in step~5-C.I of Algorithm~\ref{alg:NSLR_base}.
	If we fix $\{\mathcal{G}_i^j\}$ for each iteration, we have the chance to obtain the theoretical convergence analysis as in Section \ref{the:conv}.
	However, the empirical experience in the previous works~\cite{chang2017hyper,Yuan_PAMI_2019} has proved that
	the re-matching of non-local similar groups can boost the restoration performance.
	In this paper, we also perform such a re-matching step on the reduced image $\bar{\mathcal{M}}_i$.
	As in step 5-C.I of Algorithm~\ref{alg:NSLR_base},
	we utilize $k$-NN~\cite{gu2014weighted} for the $j$-th exemplar patch on $\bar{\mathcal{M}}_i$. The explanation of this empirical enhancement is that as the iterations, the noise level of $\bar{\mathcal{M}}_i$ decreases, and therefore, the accuracy of non-local similar groups via $k$-NN also gradually increases.

In \eqref{eq:proxavg2}, $\Px{\lambda_0 r}{\mathcal{A}} = \arg\min_{\mathcal{A}} \frac{1}{2} \| \mathcal{A}_i^j - \mathcal{A} \|_F^2 + \lambda_0 r(\mathcal{A})$
is known as the proximal operation in the optimization literature \cite{parikh2014proximal}.
Instead,
simple-closed form exists for \eqref{eq:proxavg2}.
We reshape $\mathcal{A}_i^j$ to the matrix along non-local dimension, and adopt a low-rank regularizer,
i.e., WNNM~\cite{gu2014weighted},
here in this paper.
\eqref{eq:proxavg2} can be computed by a singular value decomposition (SVD) on the matrix version of patch group $\mathcal{A}_i^j$.
In \eqref{eq:proxav3}, the denoised group tensors are denoted as ${\mathcal{M}^j}$,
and they can be directly used to reconstruct reduced image $\mathcal{M}_i$. The output of the recovered HSI of $i$-th iteration is $\mathcal{X}_i=\mathcal{M}_i\times_3\mathbf{A}_i$.

The performance superiority of our paradigm is guaranteed by state-of-the-art non-local spatial denoising \eqref{eq:proxavg2}, meanwhile, the computation complexity is controlled by the low-rank orthogonal basis exploration~\eqref{eq:sub:a_new}.
In this paper, we focus on the design of the proposed paradigm,
since the spectral low-rank property has been explored by \eqref{eq:sub:a_new},
we simply use matrix based WNNM \cite{gu2014weighted} to denoise each non-local patch group $\left[ \bar{\mathcal{M}}_i \right]_{\mathcal{G}_i^j}$.
The better choice of the reduced image denoiser is explored in Tab.~\ref{tab:WNNMvsWSNM}.
%However, , and choose the classical method WNNM for fair comparison~\cite{chang2017hyper,Yuan_PAMI_2019}.

\subsubsection*{Step~6-D.I: Noise re-estimation}

When we adopt WNNM~\cite{gu2014weighted} to solve the subproblem of \eqref{eq:proxavg2},
we need to estimate the parameter $\lambda_0 = \frac{\sigma_i^2}{2}$ with ${\sigma_i^2}$ standing for the noise level of $\bar{\mathcal{M}}_i$,
which changes during the iterations. Please note that
many non-local denoising methods assume the noise in $\bar{\mathcal{M}}_i$ follows a univariate Gaussian distribution \cite{rudin1992nonlinear,wang2008new,mairal2009non,dong2013nonlocal,gu2014weighted}.
If such assumption fails,
the resulting performance can deteriorate substantially.
Here,
we have the following Proposition~\ref{pr:nslevel}.
Therefore,
the noise distribution is preserved from $\mathcal{Y}$ to $\mathcal{\bar{M}}_i$,
which enables us to use the existing spatial denoising methods.

\begin{proposition}
	\label{pr:nslevel}
	Assume the noisy HSI $\mathcal{Z}$ is from \eqref{eq:overall_L},
	then the noise on the reduced image $\mathcal{Z} \times_3 \mathbf{P}^{\top}$,
	where $\mathbf{P}^{\top} \mathbf{P} = \mathbf{I}$,
	still follows Gaussian distribution with zero mean and variance $\sigma_0^2$.
\end{proposition}

From Proposition~\ref{pr:nslevel},
we know the noise level of $\bar{\mathcal{M}}_i$ is the same as that of $\mathcal{X}_i$;
thus, we propose to estimate it via
\begin{equation}
\sigma_i
=\gamma \times\sqrt{\vert \sigma_0^2 - \Vert  \mathcal{X}_i - \mathcal{X}_0 \Vert_F^2 / (MNB)},
\label{eq:noise-estimation}
\end{equation}
where $\gamma$ is the scaling factor controlling the re-estimation of noise variance.
%and $\text{mean}(\cdot)$ stands for the average of the tensor elements.

\subsubsection*{Step~6-D.II: Rank adaptation}

The rank of subspace $\mathbf{A}$ leads to a trade-off between noise removal ability (lower ranks) and image preservation ability (higher ranks). In Algorithm \ref{alg:NSLR_base}, $\mathbf{A}$ is refined during the iteration, to adaptively adjust the estimated rank to boost the restoration performance.
%The above approaches together make the algorithm very efficient.
At the beginning of Algorithm~\ref{alg:NSLR_base},
because the recovered image $\mathcal{X}_{i-1} = \mathcal{M}_{i-1} \times_3 \mathbf{A}_{i-1}$ is of low quality,
the latent input image $\mathcal{Z}_i$ obtained by \eqref{eq:sub1} is also of low quality.
As shown in \eqref{eq:diclearning},
the orthogonal basis is substantially influenced by the low quality of latent input image $\mathcal{Z}_i$.
Fortunately,
after the spectral denoising
%\footnote{{\color{blue}+++ need changes.}}
\eqref{eq:diclearning} and spatial denoising \eqref{eq:NLdenoising},
the quality of $\mathcal{X}_{i} = \mathcal{M}_{i} \times_3 \mathbf{A}_{i}$ is substantially improved.
As a result, in the next stage,
the quality of $\mathcal{Z}_{i+1}$ is improved,
resulting in a better estimation of the orthogonal basis via \eqref{eq:diclearning} and the reduced image via \eqref{eq:NLdenoising}.

We initialize $K$ as a small value by HySime~\cite{BioucasTGRS2008} (in step~1).
When the latent input image $\mathcal{X}$ is of low quality,
	the estimated $K$ can be small to get rid of the noise and achieve the satisfied restored results. From another perspective, real HSIs have approximate spectral low-rank property and the noisy-free HSIs are of full-rank. It can be concluded that a smaller value $K$ results in the details missing in the HSI restoration.
From the above discussion,
%Fortunately, the larger singular values obtained from the input latent image are less contaminated by the noise,
%and help to keep noise out of the reduced image.
as the iteration,
we are motivated to increase $K$ by
\begin{equation}
K = \text{min}(K + \delta \times i, B),
\label{delta}
\end{equation}
where stepsize $\delta$ and $B$ are constant values. $\delta$ is the spectral rank estimation gap between the input $\mathcal{Z}_i$ of nearby iterations, and $B$ is the spectral size of image $\mathcal{X}$. As the iterations, the noise variance decreases, and we need a larger value $K$ to capture more details of the image.
Therefore, $\mathbf{A}_{i+1}$ has the ability to capture more useful information with the number of iterations.
%\footnote{{\color{red}+++ do you need to assume the max $K$?
%	if many iterations are needed, $K$ can be higher than the tensor size.}}
As we increase the dimension $K$ of $\mathbf{A}_{i+1}$ in the next iteration, we will meet the challenge to optimize \eqref{eq:sub:a_new}. We introduce two empirical methods to optimize \eqref{eq:sub:a_new} via \eqref{eq:diclearning} and \eqref{eq:sub:a_new_r3}, respectively. Furthermore, the optimization of $\mathbf{A}$ via the above two empirical methods will produce almost the same HSI restoration results as analyzed in Section~\ref{Empirical:analysis}.

\subsection{Convergence Analysis}
\label{the:conv}
In this subsection, we fix the ideal cases of noise variance $\sigma_i^2$, dimension $K$, and non-local groups $\left\{\mathcal{G}^j\right\}$ for each iteration.
To analyze the convergence behavior of Algorithm~\ref{alg:NSLR_base},
the main problem here is that we have many overlapped regularizers in \eqref{eq:overall_L},
which are approximated by \eqref{eq:proxav1}-\eqref{eq:proxav3}.
To address this issue,
we will make use of the proximal average theory \cite{bauschke2008proximal,yu2013better,yao2019efficient} as mathematics tools.
Note that the tool is also new to the hyper-spectral image literature,
as previous analysis is mostly based on ADMM~\cite{Yuan_PAMI_2019}.
Thus,
they ignore the approximation problem on non-local patch groups.
First,
we make Assumption~\ref{ass:r}
on the regularizer $r$.
Assumption~\ref{ass:r}
is very general,
which covers the TV \cite{HE2016TGRS}
and
weighted nuclear norm \cite{gu2014weighted} regularization.

\begin{assumption}[\cite{tao2005dc}]
\label{ass:r}
The regularizer $r$ can be decomposed as
$r = r_1 - r_2$ where $r_1$ and $r_2$ are convex functions.
\end{assumption}

\begin{proposition} \label{pr:prox}
There exists a function $\bar{r}$
such that $\mathcal{M}^* = \text{prox}_{\bar{r}}\left( \bar{\mathcal{M}}_i \right)$
where $\mathcal{M}^*$ is in \eqref{eq:proxav3} and $\bar{\mathcal{M}}_i$ is in \eqref{eq:NLdenoising}.
\end{proposition}

From Proposition~\ref{pr:prox},
we can see Algorithm~\ref{alg:NSLR_base} is optimizing another objective $\bar{F}$,
which is defined as
\begin{align*}
\bar{F}(\mathbf{A}, \mathcal{M}, \mathcal{Z})
=
& \frac{1}{2}
\left\| \mathcal{Y} \! - \! h( \mathcal{Z} ) \right\|_F^2
+ \lambda \bar{r}(\mathcal{M})
\\
& + \frac{\mu}{2} \left\| \mathcal{Z}  - \mathcal{M} \times_3 \mathbf{A} \right\|_F^2,
\;\text{s.t.}\;
\mathbf{A}^{\top} \mathbf{A} = \mathbf{I}.
\end{align*}
The approximation is guaranteed in Proposition~\ref{pr:approx},
where the constant error is caused by using \eqref{eq:proxav1}-\eqref{eq:proxav3} to generate \eqref{eq:NLdenoising}.

\begin{proposition}\label{pr:approx}
$0 \le \min F - \min \bar{F} \le \lambda_0 c$ where $c$ is a constant.
\end{proposition}

Subsequently,
the convergence of Algorithm~\ref{alg:NSLR_base} is justified in below Remark~\ref{thm:conv}.

\begin{remark} \label{thm:conv}
Assume
$\bar{F}$ is coercive,
step C.I (re-matching) and D.I-II (adaptation) are not performed,
then each sequence $\{ \mathbf{A}_i \}$,
$\{\mathcal{M}_i \}$,
$\{ \mathcal{Z}_i \}$ generated from Algorithm~\ref{alg:NSLR_base} has at least one limit points,
and all limit points are also critical points of $\bar{F}$.
\end{remark}

%As $\lambda_0 \rightarrow 0$ the approximation error will finally reduce to zero,
%and Algorithm~\ref{alg:NSLR_base} will exactly solves \eqref{eq:overall_L}.
%However,
%the non-local regularization effect disappears
%and the recover performance becomes bad.
%Thus, $\lambda_0$ is a trade-off parameter. We define $\lambda_0 = \nicefrac{\sigma_i^2}{2}$ which is estimated via \eqref{eq:noise-estimation}.

\subsection{The Importance of Iterative Refinement}

%\footnote{{\color{blue}+++ the old title here is not proper now,
%	we will have another convergence analysis subsection.}}
Because denoising, compressed HSI reconstruction and inpainting can be formulated as the same optimization model~\eqref{eq:overall_L} with different degradation operators $h$, we take HSI denoising as an representative example to look into \eqref{eq:overall_L}, and determine why NGmeet should perform better than all previous spectral low-rank methods~\cite{zhuang2018fast}.
First, we initialize $\mathcal{Z}_1=\mathcal{Y}$. Recall that in~\eqref{eq:overall_L}, our model tries to exploit the spectral low-rank property and decompose the noisy latent input $\mathcal{Z}_1$ into a coarse spectral
low-rank orthogonal basis $\mathbf{A}$ and reduced image $\mathcal{M}$. Specifically,
the $i$-th column of $\mathbf{A}$,
denoted as $\mathbf{A}(:,i)$,
is regarded as the $i$-th signature of HSI,
and the corresponding coefficient image $\mathcal{M}(:,:,i)$ is regarded as the abundance map. However,
we model the spatial and spectral low-rank properties simultaneously,
which enables an iterative refinement of the orthogonal basis matrix $\mathbf{A}$.
To demonstrate the advantage of our model,
we calculated the orthogonal basis $\mathbf{A}_1$ and reduced image $\bar{\mathcal{M}}$ from noisy WDC with noise variance 50. The reference ${\mathbf{A}}$ and ${\mathcal{M}}$ are from the original clean WDC.

\begin{figure}[ht]
	\centering
	\includegraphics[width= 0.9 \linewidth]{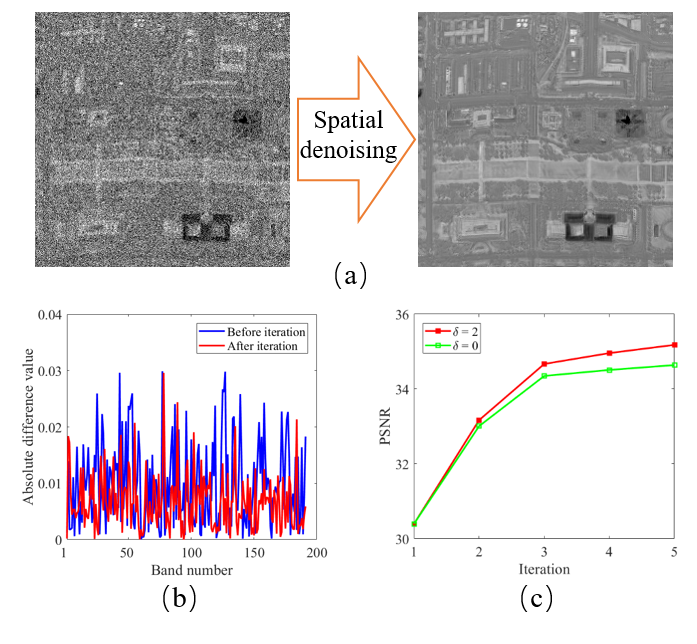}
	%%\vspace{-15px}
	\caption{(a) displays the coefficient images $\bar{\mathcal{M}}(:,:,4)$ before and after spatial denoising \eqref{eq:NLdenoising}.
		(b) displays the absolute difference signature between $\mathbf{A}(:,4)$ and the reference before and after iteration.
        (c) displays the PSNR values of $\mathcal{X}_{iter}$ with $\delta$ equal to $0$ and $2$ in \eqref{delta}.
        The test dataset is WDC with noise variance 50.}
	\label{fig:Motivation1}
\end{figure}

We firstly illustrate the advantage of our iterative paradigm model for orthogonal basis
matrix and reduced image rectification.
Fig.~\ref{fig:Motivation1}(a) presents the coefficient (or reduced) image before and after spatial denoising \eqref{eq:NLdenoising}.
These results can also be obtained by a previous method~\cite{zhuang2018fast}.
Fig.~\ref{fig:Motivation1}(b) presents the absolute difference signature between the reference $\mathbf{A}$ and $\mathbf{A}_1$ ($\mathbf{A}_2$) obtained by our paradigm. Typically, the smaller absolute difference signature means a clearer orthogonal basis. Fig.~\ref{fig:Motivation1}(b) shows that the orthogonal basis $\mathbf{A}$ is refined by our proposed model iteratively, however ignored in~\cite{zhuang2018fast}.
Subsequently, we illustrate the advantage of our rank adaptation strategy \eqref{delta}.
$\delta = 0$ in \eqref{delta} means the fixed rank in the paradigm iterations. The rank adaptation with $\delta>0$ helps the model to more precisely capture the detailed image information, as displayed in Fig.~\ref{fig:Motivation1}(c).

\section{Application Examples}
\label{sec:application}

In this section, we implement the details of proposed NGmeet for three HSI restoration problems:
denoising, compressed HSI reconstruction, and inpainting.

\subsection{HSI denoising}
\label{sec:denoise}

HSIs have been proved useful in different applications such as remote sensing~\cite{Stein2002}, medical diagnosis~\cite{Medicalhyper2014},
and face recognition~\cite{HyperfaceTPAMI2003,facehyperTIP2015}. However, during the imaging process, the HSIs are often corrupted by instrumental noise.
Therefore, HSI denoising is necessary for subsequent applications.
The purpose of HSI denoising is to obtain a clean image from the noisy image.
For the denoising task using the proposed NGmeet, the degradation operator $h$ is the identity operator,
and $\mathcal{N}$ stands for Gaussian distributed noise. The objective function now becomes
\begin{align}
\label{eq:overall_denosing}
\left\lbrace \mathbf{A}^*, \mathcal{M}^*, \mathcal{Z}^* \right\rbrace =
& \underset{\mathbf{A}, \mathcal{M}, \mathcal{Z}}{\arg\min}
\frac{1}{2}
\left\| \mathcal{Y} \! - \! \mathcal{Z}  \right\|_F^2
+ \lambda \left\| \mathcal{M} \right\|_{\text{NL}}
\\
& + \frac{\mu}{2} \left\| \mathcal{Z} - \mathcal{M} \times_3 \mathbf{A}\right\|_F^2,
\;\text{s.t.}\;
\mathbf{A}^{\top} \mathbf{A} = \mathbf{I},
\notag
\end{align}
Specifically, the noise variance is known in advance.
We adopt a multiple regression theory-based approach~\cite{BioucasTGRS2008} to estimate the noise variance for the real HSI dataset.
The update of the latent input image $\mathcal{Z}$ is
\begin{align}
%\mathcal{Z}_i
%=\frac{1}{1+\mu} \mathcal{Y}+\frac{\mu}{1+\mu} \mathcal{M}_{i-1}\!\times_3\! \mathbf{A}_{i-1}.
\mathcal{Z}_i
%\times_3 (\mathbf{I}+ \mu \mathbf{A} \mathbf{A}^{\top})
= \frac{1}{1+\mu}
\left( \mathcal{Y}+\mu \mathcal{M}_{i-1}\!\times_3\! \mathbf{A}_{i-1}
\right) ,
\label{eq:opt:sub1_v2}
\end{align}
By carefully choosing parameter $\mu$ and initializing $\mathcal{X}_0=0, \mathcal{Z}_1=\mathcal{Y}$, the optimization of \eqref{eq:opt:sub1_v2} becomes a special case of iteration strategy proposed in our previous paper~\cite{he2018non}.

\subsection{Compressed HSI reconstruction}
\label{sec:compress}

High spectral resolution is beneficial for the applications of HSI but heavily increases the burden of data storage and transmission~\cite{Bioucas2012jstars,wang2016adaptive,zhang2018cluster}.
The aim of compressed HSI reconstruction is to compress the HSI via a small number of measurements in the encoder stage, so that the compressed data can be reconstructed in the decoder stage~\cite{zhang2018cluster}.
In this paper, we focus on the decoder stage, which recovers the HSI from the compressed data. We perform two kinds of compressed measurements in the encoder stage to evaluate the efficiency of the proposed NGmeet reconstruction method.
First, the same as in \cite{zhang2018cluster,peng2018enhanced},
we assume that the original HSI is available, and the random permuted Hadamard transform operator is adopted to compress (decoder) the image. Second, as introduced in \cite{arce2013compressive,Yuan_PAMI_2019},
a compressive sensor is used to obtain a coded compressed image with a prior designed measurement operator, which is called compressive HSI imaging.
In both cases, we can assume that the sampling measurement operator $h$ is known in advance, and our proposed NGmeet method can be successfully used to reconstruct the HSI, which is formulated as:
\begin{align}
\label{eq:overall_CS}
\left\lbrace \mathbf{A}^*, \mathcal{M}^*, \mathcal{Z}^* \right\rbrace =
& \underset{\mathbf{A}, \mathcal{M}, \mathcal{Z}}{\arg\min}
\frac{1}{2}
\left\| \mathcal{Y} \! - h( \mathcal{Z})  \right\|_F^2
+ \lambda \left\| \mathcal{M} \right\|_{\text{NL}}
\\
& + \frac{\mu}{2} \left\| \mathcal{Z} - \mathcal{M} \times_3 \mathbf{A} \right\|_F^2,
\;\text{s.t.}\;
\mathbf{A}^{\top} \mathbf{A} = \mathbf{I}.
\notag
\end{align}

To enhance consistency of the reconstructed image $\mathcal{X}$ with measured data $\mathcal{Y}$, we need to optimize \eqref{eq:sub1}, which is equivalent to the following problem:
\begin{align}
h^{\star} h(\mathcal{Z}_i)+\mu \mathcal{Z}_i
= {h^{\star}}\mathcal{Y}+\mu {M}_{i-1}\!\times_3\! \mathbf{A}_{i-1},
\label{eq:opt:CS}
\end{align}
where $h^{\star}$ is the adjoint of $h$~\cite{peng2018enhanced,Yuan_PAMI_2019}.
The optimization of \eqref{eq:opt:CS} can be efficiently solved by the preconditioned conjugate gradient method~\cite{peng2018enhanced,Yuan_PAMI_2019}.

In compressed HSI reconstruction, the HSI is free from Gaussian noise, but suffers from other kinds of degradations.
We initialize the variance of the Gaussian noise to a small value to ensure the success of  NGmeet.
Other parameters, such as $\lambda$, $\mu$, and $\gamma$ are set the same values used in the HSI denoising task.
Typically, the measurement operator $h$ is poorly conditioned,
and a good initialization of $\mathcal{X}_0=\mathcal{M}_{0} \times_3 \mathbf{A}_{0}$ is necessary to predict a satisfactory latent input image $\mathcal{Z}$ via \eqref{eq:sub1}.
We adopted the initialization method as introduced in \cite{xue2019nonlocal} for the random measurement compressed case and the initialization method introduced in \cite{Yuan_PAMI_2019} for the compressive HSI imaging case.

\subsection{HSI inpainting}
\label{sec:impaint}

Because of sensor failure and poor weather conditions, remote sensing images often suffer from missing information, such as stripes and clouds~\cite{zhuang2017hyperspectral,he2019total}, that substantially influence the subsequent applications.
The aim of HSI inpainting is to predict a clean image from the observed HSI with missing information~\cite{xie2018tensor,yao2019efficient}.
In the inpainting case, $h$ is the sampling operator and we adopt $\Omega$ to represent the locations of the sampling pixels. The objective function is
\begin{align}
\label{eq:overall_inpainting}
\!\!\!
\left\lbrace \mathbf{A}^*, \mathcal{M}^*, \mathcal{Z}^* \right\rbrace
& \! = \! \underset{\mathbf{A}, \mathcal{M}, \mathcal{Z}}{\arg\min}
\frac{1}{2}
\left\| \mathcal{Y} \! - \mathcal{P}_{\Omega}( \mathcal{Z})  \right\|_F^2
\! + \! \lambda \left\| \mathcal{M} \right\|_{\text{NL}},
\\
& + \frac{\mu}{2} \left\| \mathcal{Z} - \mathcal{M} \times_3 \mathbf{A} \right\|_F^2,
\;\text{s.t.}\;
\mathbf{A}^{\top} \mathbf{A} = \mathbf{I},
\notag
\end{align}
where $\mathcal{P}_{\Omega}$ is the projection to the observed $\Omega$.
Then, the optimization of the latent input image $\mathcal {Z}$ via \eqref{eq:sub1} becomes
\begin{equation*}
(\mathcal{Z}_i)_{(i,j,k)}=
\begin{cases}
\mathcal{Y}_{(i,j,k)}
&
(i,j,k) \in \Omega \\
(\mathcal{M}_{i-1}\!\times_3\! \mathbf{A}_{i-1})_{(i,j,k)}
&
(i,j,k) \notin \Omega\\
\end{cases}.
\end{equation*}
Finally,
we initialize
parameters $\lambda$, $\mu$, and $\gamma$ have the same values as those in the HSI denoising task;
and warm start $\mathcal{X}_0=\mathcal{M}_{0}\!\times_3\! \mathbf{A}_{0}$ as \cite{xie2018tensor}.

\subsection{Time complexity analysis}

We compare the time complexity of NGmeet with other state-of-the-art non-local methods~\cite{chang2017hyper,xie2017kronecker}.
Taking the application in Section~\ref{sec:denoise} as an example,
the main time complexity of each iteration in Algorithm 1 includes stage A---SVD $(\mathcal{O}(MNB^2))$,
and stage B---non-local low-rank denoising of each $\mathcal{G}^j$ $\mathcal{O}(Tn^2K p^2)$.
For different application tasks, the optimization of \eqref{eq:sub1} is different but cheap to compute,
and therefore, is omitted from the time complexity analysis.
Tab.~\ref{tab:time_com} presents the time complexity comparison between NGmeet with those of other non-local HSI denoising method.
Here, LLRT and KBR only need stage B to perform denoising.
As the results show,  NGmeet costs an additional $\mathcal{O}(MNB^2)$ complexity in stage A,
however, it will be at least $B/K$ times faster in stage B.

\begin{table}[ht]
\caption{Complexity comparison of each iteration between proposed NGmeet and state-of-the-art non-local based methods.
	$\mathcal{G}^j \in  \mathbb{R}^{n\times{}n\times{}K\times{}p}$, where $n$ is the size of each patch and $p$ is the number of similar
	patches. $T$ is the number of $\{\mathcal{G}^j \}$ and $T_o$ is the inner iteration of KBR.}
%\vspace{-10px}
\centering
\begin{tabular}{ c | c | C{120px} }
	\hline
	                            & stage A              & stage B                                           \\ \hline
	NGmeet                      & $\mathcal{O}(MNB^2)$ & $\mathcal{O}(T n^2Kp^2)$                          \\ \hline
	LLRT \cite{chang2017hyper}  & ---                  & $\mathcal{O}(T n^2Bp^2)$                          \\ \hline
	KBR \cite{xie2017kronecker} & ---                  & $\mathcal{O}(TT_0( n^2Bp(n^2+B+p)+ n^6+B^3+p^3))$ \\ \hline
\end{tabular}
\label{tab:time_com}
\end{table}

\section{Experiments}

\begin{table*}[t]
	\centering
	\footnotesize
	\caption{Quantitative comparison of different algorithms in the simulated HSI denoising experiments.
		The PSNR is in dB, and best results are in bold.}
	%\vspace{-10px}
	\begin{tabular}{c|c|c|C{27px}|C{27px}|C{27px}|C{27px}|C{27px}|C{27px}||C{27px}|C{27px}|C{27px}||C{27px}}
		\hline
		      &          &       &             \multicolumn{6}{c||}{spectral low-rank}             & \multicolumn{3}{c||}{spatial non-local similarity} &                \\ \hline
		Image & $\sigma$ & Index & LRTA  & LRTV  &    MTS-NMF    & NAIL-RMA & PARA-FAC & Fast-HyDe &  TDL  &  KBR  &                LLRT                &     NGmeet     \\ \hline
		      &          & PSNR  & 44.12 & 41.47 &     44.27     &  28.51   &  38.01   &   46.72   & 45.58 & 46.20 &               47.14                & \textbf{47.89} \\ \cline{3-13}
		CAVE  &    10    & SSIM  & 0.969 & 0.949 &     0.972     &  0.941   &  0.921   &   0.985   & 0.983 & 0.980 &               0.989                & \textbf{0.990} \\ \cline{3-13}
		      &          &  SAM  & 7.90  & 16.54 &     8.49      &  14.52   &  13.86   &   6.62    & 6.07  & 8.94  &           \textbf{4.65}            &      4.71      \\ \cline{2-13}
		      &          & PSNR  & 38.68 & 35.32 &     37.18     &  35.11   &  37.58   &   41.21   & 39.67 & 41.52 &               42.53                & \textbf{43.15} \\ \cline{3-13}
		      &    30    & SSIM  & 0.913 & 0.818 &     0.855     &  0.775   &  0.888   &   0.945   & 0.942 & 0.942 &           \textbf{0.974}           &     0.972      \\ \cline{3-13}
		      &          &  SAM  & 12.86 & 33.32 &     14.97     &  32.43   &  17.37   &   14.06   & 12.54 & 19.43 &                8.23                & \textbf{7.43}  \\ \cline{2-13}
		      &          & PSNR  & 35.49 & 32.27 &     33.40     &  32.11   &  30.06   &   38.05   & 36.51 & 39.41 &               40.09                & \textbf{40.43} \\ \cline{3-13}
		      &    50    & SSIM  & 0.858 & 0.719 &     0.730     &  0.638   &  0.571   &   0.889   & 0.888 & 0.922 &               0.950                & \textbf{0.950} \\ \cline{3-13}
		      &          &  SAM  & 16.53 & 43.65 &     19.06     &  22.85   &  38.35   &   20.08   & 18.23 & 21.31 &               11.48                & \textbf{9.77}  \\ \cline{2-13}
		      &          & PSNR  & 31.21 & 27.97 &     27.96     &  27.90   &  24.29   &   33.41   & 31.90 & 33.78 &               36.25                & \textbf{37.19} \\ \cline{3-13}
		      &   100    & SSIM  & 0.735 & 0.529 &     0.493     &  0.453   &  0.256   &   0.746   & 0.734 & 0.851 &               0.910                & \textbf{0.926} \\ \cline{3-13}
		      &          &  SAM  & 22.67 & 54.85 &     26.33     &  55.66   &  51.83   &   30.72   & 28.51 & 26.41 &               18.17                & \textbf{16.22} \\ \hline
		      &          & PSNR  & 38.49 & 38.71 &     40.64     &  41.46   &  33.39   &   42.22   & 41.46 & 40.09 &               41.95                & \textbf{43.18} \\ \cline{3-13}
		 PaC  &    10    & SSIM  & 0.975 & 0.979 &     0.988     &  0.987   &  0.866   &   0.990   & 0.988 & 0.984 &               0.989                & \textbf{0.992} \\ \cline{3-13}
		      &          &  SAM  & 4.90  & 3.29  &     2.76      &   3.46   &   9.05   &   2.99    & 3.06  & 2.86  &                2.75                & \textbf{2.60}  \\ \cline{2-13}
		      &          & PSNR  & 32.07 & 32.76 &     35.45     &  34.17   &  30.92   &   35.98   & 34.43 & 34.39 &               35.04                & \textbf{36.99} \\ \cline{3-13}
		      &    30    & SSIM  & 0.908 & 0.920 &     0.958     &  0.941   &  0.845   &   0.962   & 0.949 & 0.947 &               0.957                & \textbf{0.972} \\ \cline{3-13}
		      &          &  SAM  & 7.88  & 5.76  & \textbf{4.17} &   6.54   &   9.28   &   5.09    & 5.11  & 4.28  &                4.86                &      4.26      \\ \cline{2-13}
		      &          & PSNR  & 29.11 & 29.45 &     32.51     &  30.71   &  29.24   &   33.32   & 31.31 & 31.05 &               32.00                & \textbf{34.42} \\ \cline{3-13}
		      &    50    & SSIM  & 0.836 & 0.850 &     0.921     &  0.886   &  0.846   &   0.936   & 0.904 & 0.892 &               0.918                & \textbf{0.949} \\ \cline{3-13}
		      &          &  SAM  & 9.20  & 8.60  &     5.50      &   8.83   &  11.40   &   6.55    & 6.14  & 5.40  &                6.55                & \textbf{5.09}  \\ \cline{2-13}
		      &          & PSNR  & 25.13 & 26.22 &     28.17     &  25.76   &  23.68   &   29.90   & 27.49 & 27.80 &               28.63                & \textbf{30.71} \\ \cline{3-13}
		      &   100    & SSIM  & 0.655 & 0.729 &     0.808     &  0.728   &  0.598   &   0.873   & 0.789 & 0.793 &               0.833                & \textbf{0.892} \\ \cline{3-13}
		      &          &  SAM  & 10.17 & 12.76 &     8.40      &  12.93   &  20.22   &   8.68    & 7.67  & 6.95  &                7.68                & \textbf{6.80}  \\ \hline
		      &          & PSNR  & 38.94 & 36.64 &     37.26     &  42.57   &  32.38   &   43.06   & 41.83 & 40.58 &               41.89                & \textbf{43.71} \\ \cline{3-13}
		 WDC  &    10    & SSIM  & 0.974 & 0.968 &     0.975     &  0.989   &  0.914   &   0.991   & 0.989 & 0.986 &               0.990                & \textbf{0.993} \\ \cline{3-13}
		      &          &  SAM  & 5.602 & 4.653 &     4.429     &  3.637   &  8.087   &   3.070   & 3.680 & 3.090 &               3.700                & \textbf{2.830} \\ \cline{2-13}
		      &          & PSNR  & 32.91 & 32.42 &     34.65     &  35.87   &  31.56   &   37.39   & 34.84 & 34.75 &               36.30                & \textbf{37.92} \\ \cline{3-13}
		      &    30    & SSIM  & 0.917 & 0.909 &     0.953     &  0.958   &  0.898   &   0.971   & 0.953 & 0.951 &               0.967                & \textbf{0.975} \\ \cline{3-13}
		      &          &  SAM  & 8.331 & 5.991 &     5.557     &  7.011   &  9.009   &   5.140   & 6.400 & 5.240 &               5.460                & \textbf{4.641} \\ \cline{2-13}
		      &          & PSNR  & 30.35 & 30.12 &     32.49     &  32.56   &  29.49   &   34.61   & 31.89 & 31.61 &               33.48                & \textbf{35.14} \\ \cline{3-13}
		      &    50    & SSIM  & 0.864 & 0.849 &     0.922     &  0.919   &  0.837   &   0.948   & 0.910 & 0.900 &               0.938                & \textbf{0.956} \\ \cline{3-13}
		      &          &  SAM  & 9.43  & 7.09  &     6.71      &   9.22   &  13.64   &   6.57    & 7.94  & 6.63  &                6.43                & \textbf{5.81}  \\ \cline{2-13}
		      &          & PSNR  & 26.84 & 27.23 &     28.94     &  27.85   &  23.01   &   31.05   & 27.66 & 28.23 &               29.88                & \textbf{31.48} \\ \cline{3-13}
		      &   100    & SSIM  & 0.734 & 0.740 &     0.830     &  0.805   &  0.550   &   0.894   & 0.781 & 0.789 &               0.861                & \textbf{0.908} \\ \cline{3-13}
		      &          &  SAM  & 11.33 & 9.47  &     9.44      &  13.27   &  25.46   &   8.91    & 10.15 & 9.12  &                7.99                & \textbf{7.86}  \\ \hline
	\end{tabular}%
	\label{tab:simucase}
\end{table*}

In this section, we present experimental results for the three HSI restoration tasks,
denoising (Section~\ref{sec:den}),
compressed HSI reconstruction (Section~\ref{sec:exp:comp})
and inpainting (Section~\ref{sec:imp}).
The experiments were programmed in Matlab on a computer equipped with a CPU Core i7-7820HK and 64G memory.

\subsection{HSI denoising experiments}
\label{sec:den}

\subsubsection{Simulated-data experiments}
\label{ssec:expsim}

\noindent
\textbf{Setup.}
One multi-spectral image (MSI) from the CAVE dataset
\footnote{\url{http://www1.cs.columbia.edu/CAVE/databases/}}, and two HSI images,
\textit{i.e.}
on each from the PaC
\footnote{\url{http://www.ehu.eus/ccwintco/index.php/}}
and WDC
%\footnote{\url{https://engineering.purdue.edu/~biehl/MultiSpec/hyperspectral.html}}
\footnote{\url{https://engineering.purdue.edu/~biehl/MultiSpec/hyperspectral}}
datasets were used (Tab.~\ref{tab:imagesize}).
These images have been widely used in simulated-data studies
\cite{chang2017hyper,he2015hyperspectral,CVPR2014Meng,xie2017kronecker,zhuang2018fast}.
Following the settings in~\cite{chang2017hyper,CVPR2014Meng},
additive Gaussian noise with noise variance $\sigma_0^2$ was added to the MSI/HSIs with various values of $\sigma_0^2$ of $10, 30, 50$ and $100$. Before denoising, the whole HSIs were normalized to [0, 255].

\begin{table}[ht]
\footnotesize
\caption{Hyper-spectral images used for simulated experiments.}
%\vspace{-10px}
\centering
\begin{tabular}{ c | c | c | c }
	\hline
	                & CAVE           & PaC            & WDC            \\ \hline
	image size      & 512$\times$512 & 256$\times$256 & 256$\times$256 \\ \hline
	number of bands & 31             & 80             & 191            \\ \hline
\end{tabular}
\label{tab:imagesize}
\end{table}

\begin{figure*}[t]
\centering
\includegraphics[width=0.85\linewidth]{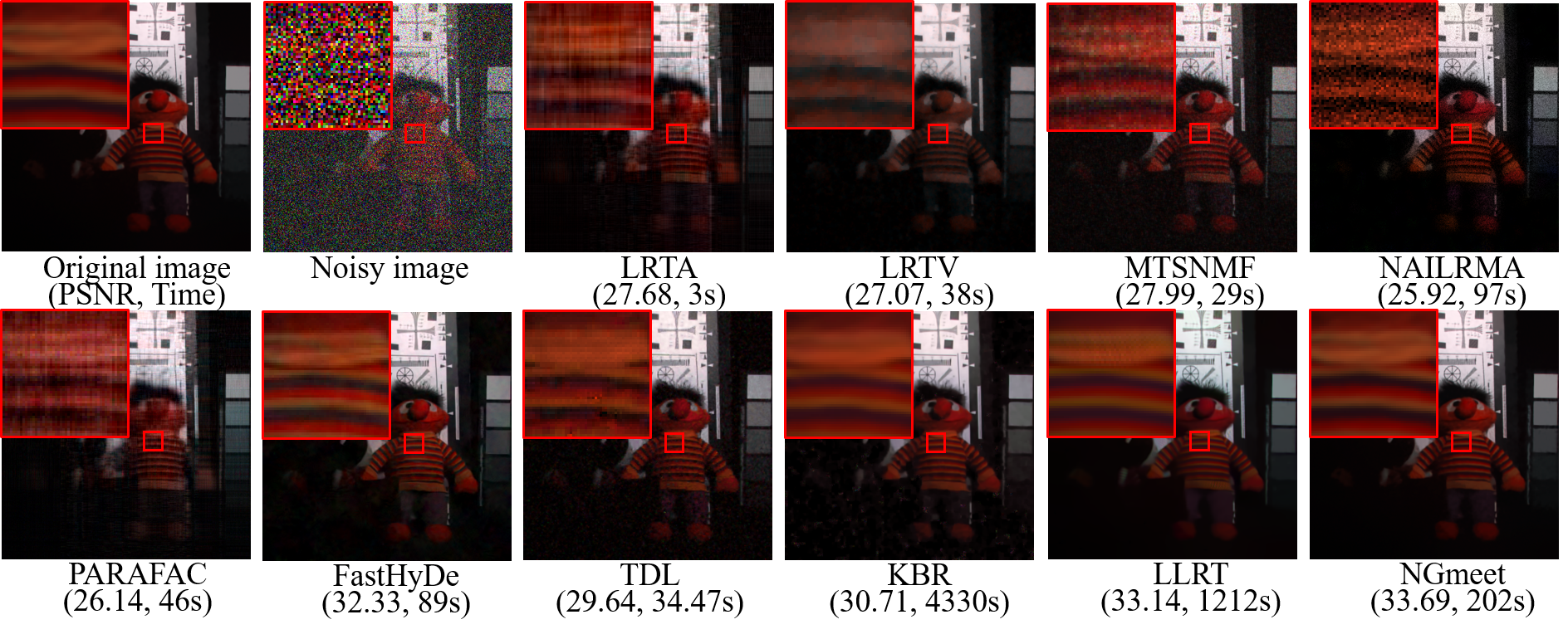}
%\vspace{-15px}
\caption{Denoising results on the CAVE-toy image with the noise variance 100. The color image is composed of bands 31, 11, and 6 for the red, green, and blue channels, respectively.}
\label{fig:simu1}
\end{figure*}

\begin{figure*}[ht]
\centering
\includegraphics[width=0.85\linewidth]{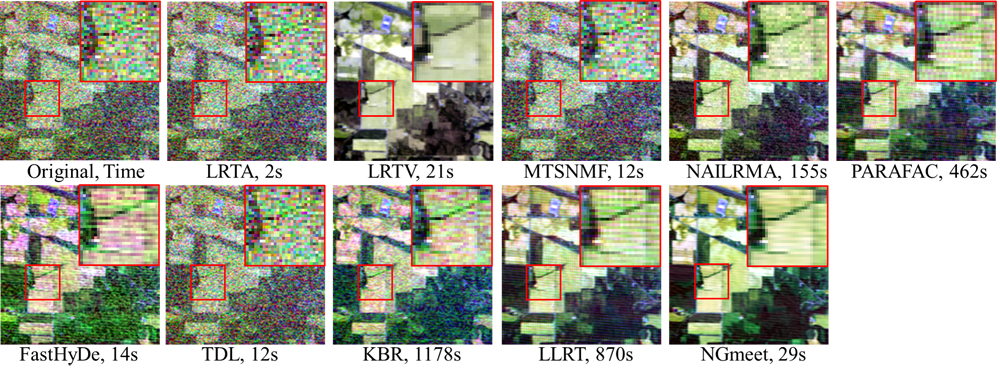}
%\vspace{-15px}
\caption{Real data experimental results on the Indian Pines dataset. The color image is composed of noisy bands 219, 109 and 1.}
\label{fig:realindian}
\end{figure*}

\begin{figure*}[ht]
\centering
\includegraphics[width=0.85\linewidth]{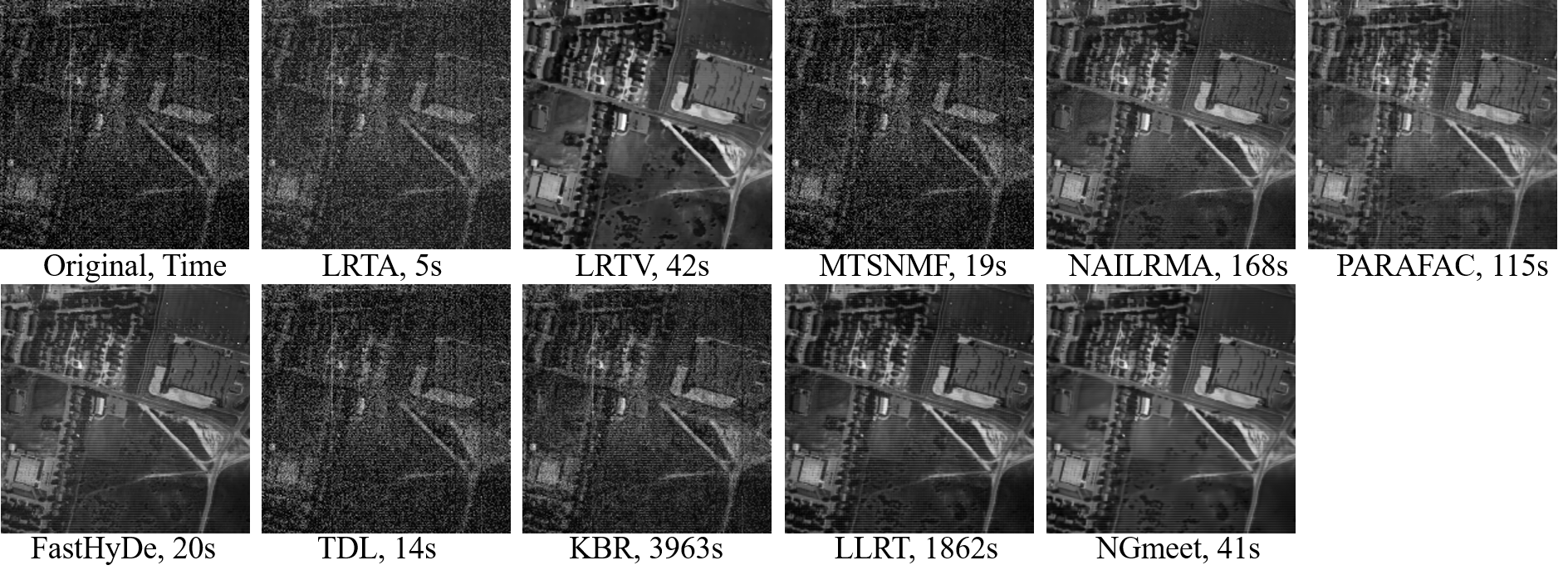}
%\vspace{-15px}
\caption{Real data experimental results on the Urban dataset of band 207.}
\label{fig:realUrban}
\end{figure*}

The following methods were compared:
\textit{spectral low-rank methods}, \textit{i.e.}
%\footnote{+++ how does hyper-parameters are set for other methods (except ours)? {\color{red} Hyper-parameters of all compared methods are set basedon authors?Ecodes or suggestions in the paper.}}
LRTA~\cite{renard2008denoising}
\footnote{\url{https://www.sandia.gov/tgkolda/TensorToolbox/}},
%\footnote{\url{https://www.sandia.gov/tgkolda/TensorToolbox/index-2.5.html}},
LRTV~\cite{HE2016TGRS}
\footnote{\url{https://sites.google.com/site/rshewei/home}},
MTSNMF~\cite{QianyuntaoTGRS2015}
\footnote{\url{http://www.cs.zju.edu.cn/people/qianyt/}},
NAILRMA~\cite{he2015hyperspectral}
%\footnote{\url{https://sites.google.com/site/rshewei/home}},
PARAFAC~\cite{liu2012denoising}
%\footnote{\url{https://www.sandia.gov/tgkolda/TensorToolbox/index-2.5.html}},
and FastHyDe~\cite{zhuang2018fast}
\footnote{\url{http://www.lx.it.pt/~bioucas/}};
%\footnote{+++ now, we do not emphasize low-rank in spatial. better change to spatial similarity methods. same for table 2.}
\textit{spatial non-local similarity methods}, \textit{i.e.}
TDL~\cite{CVPR2014Meng}
%\footnote{\url{http://gr.xjtu.edu.cn/web/dymeng/}},
KBR~\cite{xie2017kronecker}
\footnote{\url{http://gr.xjtu.edu.cn/web/dymeng/}},
LLRT~\cite{chang2017hyper}
%\footnote{\url{http://www.escience.cn/people/changyi/index.html}};
\footnote{\url{http://www.escience.cn/people/changyi/}};
and the proposed
NGmeet\footnote{\url{https://github.com/quanmingyao/NGmeet}} (Algorithm~\ref{alg:NSLR_base}),
which combines the best of the above two fields.
The hyper-parameters of all the comparison methods were set based on the authors' code or suggestions in the relevant papers.
The value of spectral dimension $K$ is the most important parameter. It was initialized by HySime~\cite{BioucasTGRS2008} and updated via \eqref{delta}.
%\footnote{+++ where can I see them in the algorithm 1? {\color{red} point out in line 9}}
Parameter $\mu$ is used to control the contribution of non-local regularization, and $\gamma$ is a scaling factor controlling the re-estimation of noise variance~\cite{dong2013nonlocal}.
We empirically set $\mu = 2,\lambda = 9$ and $\gamma = 0.5$ as recommended in~\cite{chang2017hyper}. Moreover, $\delta = 2$ in the whole experiments.

To thoroughly quantitatively evaluate the performance of the different methods, the peak signal-to-noise ratio (PSNR),
%\footnote{+++ ref for SSIM{\color{red} add}}
the structural similarity (SSIM)~\cite{SSIM2004image} and the
%\footnote{+++ should be ``spectral angle mean''? add a ref, {\color{red} Both OK.}}
spectral angle mean (SAM)~\cite{chang2017hyper,HE2016TGRS} indices were adopted.
The SAM index is used to measure the mean difference in spectral angle between the original HSI and the restored HSI.
A lower value of SAM indicates a higher similarity between the original and denoised images.

\noindent
\textbf{Quantitative comparison.}
For each noise level setting, we calculated the evaluation values of all the images from each dataset, as presented in Tab. \ref{tab:simucase}. For each image from CAVE dataset, we selected a subset of size $300 \times 300 \times 31$ for the experiments. It can be easily observed that NGmeet achieves the best results in almost all cases. Another interesting observation is that in the MSI case, the non-local based method LLRT can achieve better results than FastHyDe, which has the best result of the spectral low-rank methods, but it does the opposite in the HSI cases. This phenomenon confirms the advantage of the NL low-rank property in the MSI processing and the spectral low-rank property in the HSI processing.

\noindent
\textbf{Visual comparison.}
To further demonstrate the efficiency of NGmeet,
%\footnote{+++ u can put the other two in the appendix.}
Fig.~\ref{fig:simu1} shows the color images of the CAVE toy MSI(composed of bands 31, 11 and 6~\cite{he2015hyperspectral}) before and after denoising. The PSNR value and the computational time of each method are given under the denoised images. It can be observed that FastHyDe, LLRT, and NGmeet have huge advantage over the rest of the comparison methods. The enlarged area shows that the results of FastHyDe and LLRT contain some artifacts.
Our NGmeet method produces the best visual quality.

\begin{table}[ht]
\footnotesize
\centering
\caption{Hyperspectral images used for real data experiments.}
%\vspace{-10px}
\begin{tabular}{c | c | c}
	\hline
	                & Urban          & Indian Pines   \\ \hline
	  image size    & 200$\times$200 & 145$\times$145 \\ \hline
	number of bands & 210            & 220            \\ \hline
\end{tabular}
\label{tab:realdata}
\end{table}

\begin{table*}[!t]
  \centering
  \caption{Quantitative comparison of different algorithms in the simulated HSI compressed reconstruction experiments.
 		The best results are in bold.}
  %\vspace{-10px}
    \begin{tabular}{c|c|c|c|c|c|c|c|c|c|c} \hline
    Image & SR    & Index   & IT    & CPPCA & SpeCA & SSHBCS & CSF   & JTenRe3DTV & NTSRLR  & NGmeet \\ \hline
         &       & PSNR(dB) & 25.79 & 7.20  & 23.65 & 20.40 & 23.54 & 29.46 & 28.24 & \textbf{32.20} \\  \cline{3-11}
     Toy & 2\%   & SSIM     & 0.713 & 0.056 & 0.774 & 0.695 & 0.760 & 0.816 & 0.765 & \textbf{0.900} \\  \cline{3-11}
         &       & SAM      & 21.837 & 88.406 & 24.601 & 24.816 & 26.058 & 18.829 & 20.339 & \textbf{10.169} \\ \cline{2-11}
         &       & PSNR(dB) & 29.03 & 10.83 & 27.20 & 23.44 & 27.97 & 36.00 & 34.21 & \textbf{40.72} \\ \cline{3-11}
         & 5\%   & SSIM     & 0.826 & 0.187 & 0.825 & 0.768 & 0.849 & 0.941 & 0.904 & \textbf{0.975} \\ \cline{3-11}
         &       & SAM      & 18.948 & 61.142 & 20.786 & 25.186 & 19.373 & 9.364 & 12.679 & \textbf{4.829} \\ \cline{2-11}
         &       & PSNR(dB) & 32.62 & 21.37 & 29.68 & 24.24 & 32.43 & 41.69 & 40.52 & \textbf{47.69} \\ \cline{3-11}
         & 10\%  & SSIM     & 0.888 & 0.552 & 0.880 & 0.773 & 0.926 & 0.979 & 0.971 & \textbf{0.993} \\ \cline{3-11}
         &       & SAM      & 15.642 & 32.053 & 16.802 & 24.083 & 14.333 & 5.467 & 6.932 & \textbf{2.894} \\ \cline{2-11}
         &       & PSNR(dB) & 35.50 & 25.32 & 36.76 & 27.16 & 34.00 & 44.85 & 45.56 & \textbf{50.90} \\ \cline{3-11}
         & 15\%  & SSIM     & 0.925 & 0.649 & 0.944 & 0.857 & 0.945 & 0.988 & 0.989 & \textbf{0.996} \\ \cline{3-11}
         &       & SAM      & 12.775 & 22.513 & 9.992 & 16.188 & 13.965 & 3.930 & 4.217 & \textbf{2.273} \\ \cline{2-11}
         &       & PSNR(dB) & 38.37 & 30.61 & 37.96 & 29.05 & 35.08 & 47.76 & 48.25 & \textbf{53.82} \\ \cline{3-11}
         & 20\%  & SSIM     & 0.954 & 0.837 & 0.966 & 0.847 & 0.950 & 0.993 & 0.994 & \textbf{0.998} \\ \cline{3-11}
         &       & SAM      & 9.875 & 13.961 & 7.203 & 22.588 & 12.774 & 2.917 & 3.217 & \textbf{1.630} \\ \hline
         &       & PSNR(dB) & 23.69 & 13.85 & 27.65 & 13.86 & 26.76 & 28.60 & 28.51 & \textbf{29.02} \\ \cline{3-11}
     PaC & 2\%   & SSIM     & 0.527 & 0.288 & \textbf{0.883} & 0.061 & 0.881 & 0.812 & 0.810 & 0.834 \\ \cline{3-11}
         &       & SAM      & 9.590 & 38.986 & 11.822 & 50.352 & 13.033 & 9.873 & \textbf{7.011} & 7.451 \\ \cline{2-11}
         &       & PSNR(dB) & 26.67 & 23.43 & 32.47 & 26.71 & 31.92 & 34.32 & 32.42 & \textbf{35.00} \\ \cline{3-11}
         & 5\%   & SSIM     & 0.728 & 0.740 & 0.942 & 0.886 & 0.951 & 0.942 & 0.915 & \textbf{0.953} \\ \cline{3-11}
         &       & SAM      & 8.636 & 18.614 & \textbf{4.663} & 13.815 & 6.793 & 6.941 & 5.411 & 5.065 \\ \cline{2-11}
         &       & PSNR(dB) & 30.01 & 32.35 & 41.66 & 28.71 & 38.43 & 41.39 & 37.21 & \textbf{42.96} \\ \cline{3-11}
         & 10\%  & SSIM     & 0.862 & 0.926 & 0.987 & 0.901 & 0.972 & 0.987 & 0.970 & \textbf{0.991} \\ \cline{3-11}
         &       & SAM      & 7.867 & 7.787 & 3.478 & 13.199 & \textbf{2.769} & 3.464 & 3.912 & 2.917 \\ \cline{2-11}
         &       & PSNR(dB) & 32.47 & 41.98 & 47.54 & 36.83 & 44.16 & 42.65 & 41.31 & \textbf{49.45} \\ \cline{3-11}
         & 15\%  & SSIM     & 0.919 & 0.991 & 0.997 & 0.986 & 0.995 & 0.990 & 0.988 & \textbf{0.998} \\ \cline{3-11}
         &       & SAM      & 7.070 & 2.927 & 1.699 & 3.915 & 2.152 & 3.220 & 2.863 & \textbf{1.596} \\ \cline{2-11}
         &       & PSNR(dB) & 34.68 & 42.93 & 50.55 & 41.30 & 46.91 & 45.06 & 45.56 & \textbf{52.27} \\ \cline{3-11}
         & 20\%  & SSIM     & 0.949 & 0.992 & 0.998 & 0.991 & 0.997 & 0.994 & 0.995 & \textbf{0.999} \\ \cline{3-11}
         &       & SAM      & 6.218 & 2.630 & 1.369 & 2.893 & 1.867 & 2.563 & 2.007 & \textbf{1.189} \\ \hline
         &       & PSNR(dB) & 31.38 & 21.63 & 34.90 & 23.91 & 32.15 & 30.11 & 31.73 & \textbf{35.52} \\ \cline{3-11}
     WDC & 2\%   & SSIM     & 0.707 & 0.626 & \textbf{0.903} & 0.731 & 0.781 & 0.873 & 0.720 & 0.875 \\ \cline{3-11}
         &       & SAM      & 11.728 & 26.307 & 6.168 & 19.145 & 13.128 & 10.905 & 7.856 & \textbf{6.141} \\ \cline{2-11}
         &       & PSNR(dB) & 33.61 & 32.94 & 42.96 & 24.36 & 34.10 & 38.58 & 37.77 & \textbf{44.34} \\ \cline{3-11}
         & 5\%   & SSIM     & 0.799 & 0.739 & \textbf{0.997} & 0.828 & 0.823 & 0.979 & 0.907 & 0.981 \\ \cline{3-11}
         &       & SAM      & 9.095 & 8.338 & \textbf{2.130} & 13.389 & 10.231 & 4.353 & 3.925 & 2.364 \\ \cline{2-11}
         &       & PSNR(dB) & 36.18 & 35.93 & 54.30 & 27.22 & 33.38 & 40.81 & 42.78 & \textbf{57.33} \\ \cline{3-11}
         & 10\%  & SSIM     & 0.880 & 0.775 & 0.998 & 0.892 & 0.793 & 0.987 & 0.970 & \textbf{0.999} \\ \cline{3-11}
         &       & SAM      & 6.691 & 7.251 & 0.697 & 10.939 & 11.197 & 3.588 & 2.248 & \textbf{0.646} \\ \cline{2-11}
         &       & PSNR(dB) & 38.50 & 47.49 & 55.89 & 27.76 & 38.17 & 43.46 & 46.26 & \textbf{58.83} \\ \cline{3-11}
         & 15\%  & SSIM     & 0.927 & 0.980 & \textbf{0.999} & 0.910 & 0.887 & 0.992 & 0.987 & \textbf{0.999} \\ \cline{3-11}
         &       & SAM      & 5.089 & 1.648 & 0.584 & 11.094 & 6.676 & 2.829 & 1.517 & \textbf{0.512} \\ \cline{2-11}
         &       & PSNR(dB) & 40.83 & 49.30 & 57.00 & 29.14 & 39.46 & 44.24 & 50.40 & \textbf{59.37} \\ \cline{3-11}
         & 20\%  & SSIM     & 0.956 & 0.988 & \textbf{0.999} & 0.934 & 0.900 & 0.994 & 0.995 & \textbf{0.999} \\ \cline{3-11}
         &       & SAM      & 3.878 & 1.341 & 0.460 & 10.857 & 5.897 & 2.526 & 0.961 & \textbf{0.459} \\ \hline
    \end{tabular}%
  \label{tab:simucase2}
\end{table*}%

\begin{figure*}[ht]
\centering
\includegraphics[width=0.75\linewidth]{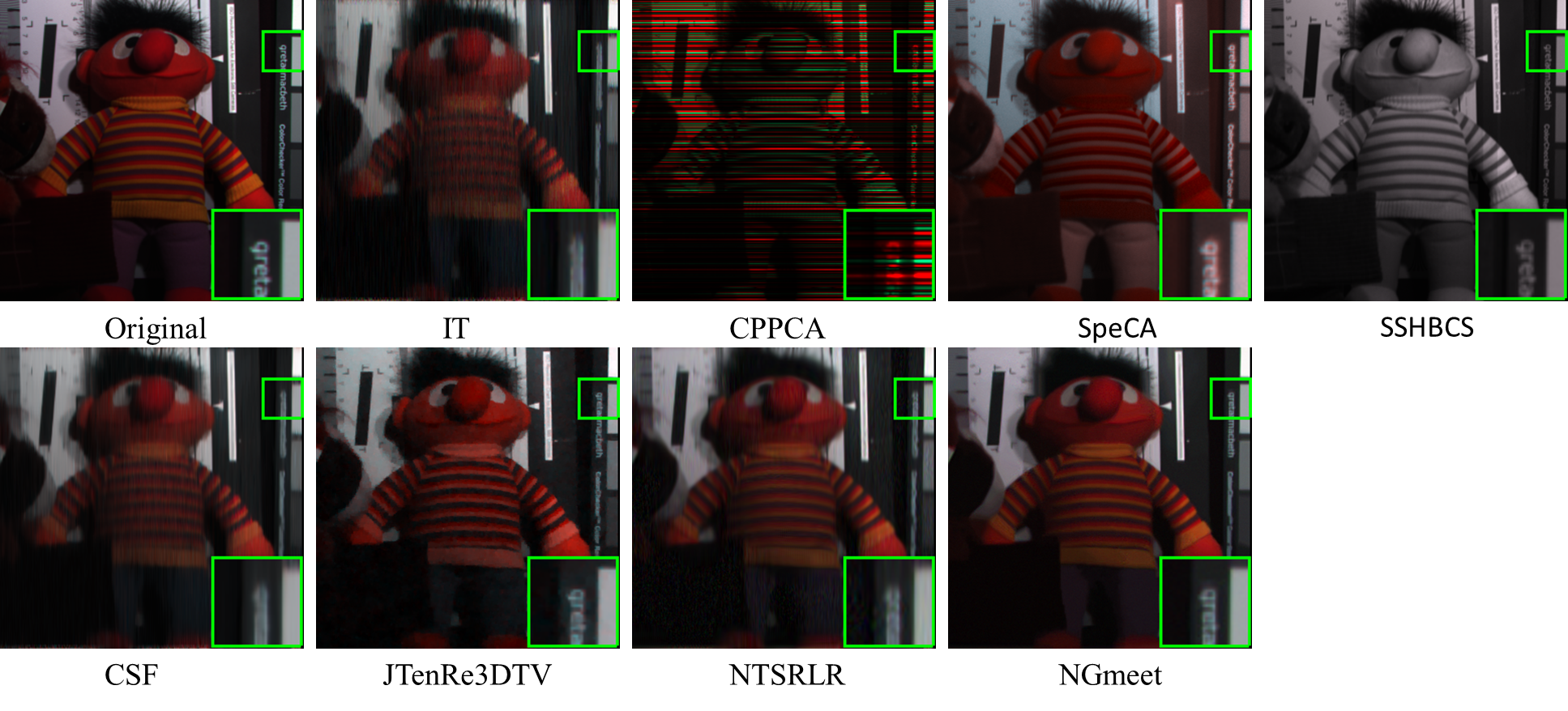}
%\vspace{-15px}
\caption{Compressed reconstruction results on the CAVE-Toy image with SR as $0.02$.}
\label{fig:compress:toy}
\end{figure*}

\begin{figure*}[ht]
\centering
\includegraphics[width=0.85\linewidth]{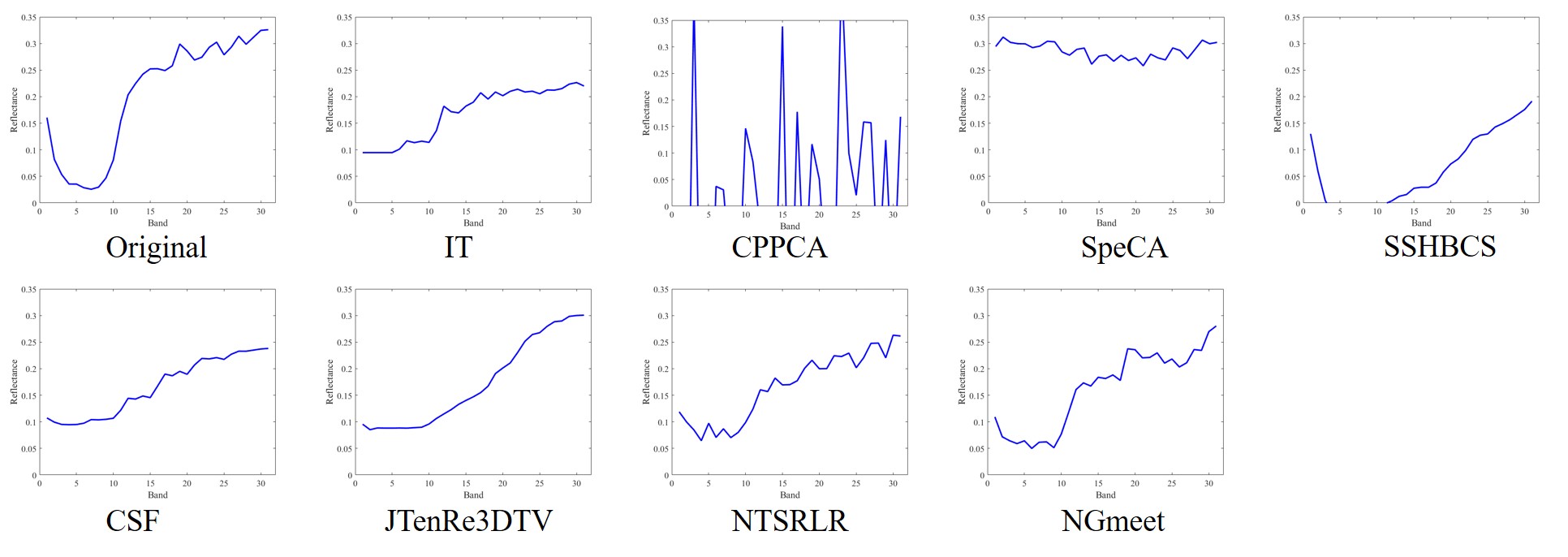}
%\vspace{-15px}
\caption{Spectral signature curves of pixel (150,150) reconstructed by different methods on CAVE-Toy with SR as $0.02$.}
\label{fig:compress:toy:sig}
\end{figure*}

%\footnote{+++ we need to add some figures to show the covergence of PSNR.}
\subsubsection{Real-data experiments}
\noindent
\textbf{Setup.}
Here, AVIRIS Indian Pines HSI
\footnote{\url{https://engineering.purdue.edu/~biehl/MultiSpec/}}
and HYDICE Urban image
\footnote{\url{http://www.tec.army.mil/hypercube}}
were adopted for the real-data experiments (Tab.~\ref{tab:realdata}). As in \cite{He2014TGRS}, 20 water absorption bands (104-108, 150-163, 220 bands) of the Indian Pines HSI were excluded for illustration, because they do not contain useful information.
%\footnote{+++ note that in our algorithm,
%    the noise-level is taken as the input.
%    how do you obtain $\sigma_0$ for real-world data sets? {\color{red}Introduced in the real dataset. By multiple regression theory-based approach}}
The noisy HSIs were also scaled to the range [0 255], and the parameters of  NGmeet was set as the same value as those in the simulated-data experiments.
In addition, multiple regression theory-based approach~\cite{BioucasTGRS2008} was adopted to estimate the initial noise variance of each HSI bands.
%\footnote{+++ please check consistency,I write ``\# of spectrum'',u use ``number of bands.''I feel the later is better. please find and replace.}
%\footnote{+++ what do PSNR in Figure 9 and 10 mean? they are real-world data,
%    how do you caculate PSNR? {\color{red} wrong, I should delate them}}

\noindent
\textbf{Visual comparison.}
%\footnote{+++ explain why we only use visual comparison here. i.e., why no SAM PSNR SSIM.}
Because clean reference images are not available for these data, we just present the real Indian Pines and Urban images before and after denoising in Figs.~\ref{fig:realindian} and~\ref{fig:realUrban}, respectively. It is clear that NGmeet can simultaneously remove the noise and keep the spectral details. LRTV produces the smoothest results. However, the color of the denoised result has large changes, indicating the loss of spectral information. The denoised results of FastHyDe and LLRT still contain stripes as shown in Fig.~\ref{fig:realindian}. Hence, although  NGmeet is designed under the assumption of Gaussian noise, it can also achieve the best results for real datasets.
%\footnote{+++ emphasize that while Guassian is assumed when designing the model, NGmeet works well for real datasets (unknown noise type) as well.}

\subsection{Compressed HSI reconstruction experiments}
\label{sec:exp:comp}

%In this section,
%we first conduct the comparison methods with a randomly permuted Hadamard transform compressed operator~\cite{wang2017compressive}. Subsequently, we also applied our proposed method to compressive HSI imaging with a fixed compressed operator.

\subsubsection{Compressed HSI reconstruction}
\noindent
\textbf{Setup.}
The MSI CAVE Toy, and two HSI images from PaC and WDC datasets were used. These datasets have also been widely used in compressed HSI reconstruction~\cite{wang2017compressive,xue2019nonlocal,zhang2018cluster}.
Following the settings in~\cite{xue2019nonlocal}, we selected a subset from the Toy image, of size $300 \times 300 \times 31$ for the experiments.
The sampling ratio (SR) varied as $2\%, 5\%, 10\%, 15\%$ and $20\%$. Before reconstruction, the MSI/HSIs were normalized to [0, 255].

The following methods were used for the comparison:
%\footnote{+++ how does hyper-parameters are set for other methods (except ours)? {\color{red} Hyper-parameters of all compared methods are set basedon authors?Ecodes or suggestions in the paper.}}
%\footnote{+++ in table 5 it is DCT.}
IT~\cite{daubechies2004iterative},
%\footnote{\url{https://www.sandia.gov/tgkolda/TensorToolbox/index-2.5.html}},
CPPCA~\cite{fowler2009compressive}\footnote{\url{http://my.ece.msstate.edu/faculty/fowler/software.html}},
SpeCA~\cite{martin2016hyperspectral}\footnote{\url{http://www.lx.it.pt/~bioucas/publications.html}},
SSHBCS~\cite{zhang2016dictionary},
%\footnote{\url{https://sites.google.com/site/leizhanghyperspectral/publications}},
CSF~\cite{zhang2018cluster}\footnote{\url{https://sites.google.com/site/leizhanghyperspectral/publications}},
JTenRe3DTV~\cite{wang2017compressive}\footnote{The code was provided by Dr. Yao Wang.},
NTSRLR~\cite{xue2019nonlocal}\footnote{The code was provided by Dr. Jize Xue.},
and the proposed
NGmeet.
The hyper-parameters of all comparison methods were set based on authors' codes or suggestions in their papers. Similar to~\cite{xue2019nonlocal},
NGmeet also adopted the IT method as the initialization.
The values of other parameters in NGmeet were the same as those in the denoising case. We implemented NTSRLR on a super-computer with 256G memory. On the WDC dataset, the super-computer was out of memory when running NTSRLR. Hence we report the results on a subimage of size $100 \times 100 \times 191$.

%The value of spectral dimension $K$ is the most import parameter, which is initialized by HySime~\cite{BioucasTGRS2008} and updated via \eqref{delta}.
%\footnote{+++ where can I see them in the algorithm 1? {\color{red} point out in line 9}}
%Parameter $\mu$ is used to control the contribution of non-local regularization, and $\gamma$ is a scaling factor controlling the re-estimation of noise variance~\cite{dong2013nonlocal}.
%We empirically set $\mu=1$, $\lambda = 0.9$ and $\gamma = 0.5$ as introduced in~\cite{chang2017hyper}, and $\delta = 2$ in the whole experiments.

\noindent
\textbf{Quantitative comparison.}
We calculated evaluation values of all the images from each dataset for each SR,
and present the results in Tab.~\ref{tab:simucase2}.
%\footnote{+++ the performance is too good here.
%	much better than baselines, need to explain more}
We can see that
NGmeet always achieves better evaluation results than the non-local similarity method NTSRLR and global-based method JTenRe3DTV.
SpeCA yields evaluation values similar to those of NGmeet for the PaU and WDC datasets.
However, it performs worse on the CAVE Toy image. CPPCA fails to reconstruct the image when the SR is low.

\noindent
\textbf{Visual comparison.}
Fig.~\ref{fig:compress:toy} shows the color images of CAVE Toy (composed of bands 31, 11 and 6) before and after compressed reconstruction via different methods.
%\footnote{{\color{red}+++ better to plot all lines in one figure.
%	this helps easier comparison. Better to plot separately, as the same of related papers}}
Fig.~\ref{fig:compress:toy:sig} illustrates the spectral signature curves of pixel (150,150) reconstructed by different methods on CAVE Toy with SR of $0.02$. SSHBCS fails to reconstruct the information of some bands, resulting in the loss of spectral information. JTenRe3DTV and NTSRLR produce some blurred details.
Overall, NGmeet achieves the best visual results.

\begin{table}[ht]
  \centering
  \caption{Quantitative comparison of different algorithms in the HSI compressive imaging reconstruction experiments.
		The PSNR is in dB.
		The best results are in bold.}
   %\vspace{-10px}
    \begin{tabular}{c|c|c|c|c|c|c}
    	\hline
    	Method & \multicolumn{2}{c|}{GAP-TV} & \multicolumn{2}{c|}{DeSCI} &    \multicolumn{2}{c}{NGmeet}    \\ \hline
    	Index  & PSNR  &        SSIM         & PSNR  &        SSIM        &      PSNR      &      SSIM       \\ \hline
    	value  & 24.66 &       0.8608        & 25.91 &       0.9094       & \textbf{27.13} & \textbf{0.9130} \\ \hline
    \end{tabular}%
  \label{tab:imaging}
\end{table}%

\begin{figure*}[ht]
\centering
\includegraphics[width=0.75\linewidth]{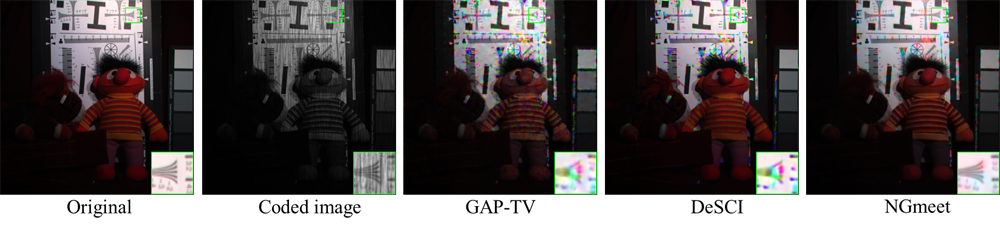}
%\vspace{-10px}
\caption{Reconstructed images of different methods from coded Toy dataset.}
\label{fig:img_toy}
\end{figure*}

\begin{figure}[ht]
	\centering
	\subfigure[PSNR v.s. band]{
		\includegraphics[width= 0.45 \linewidth]{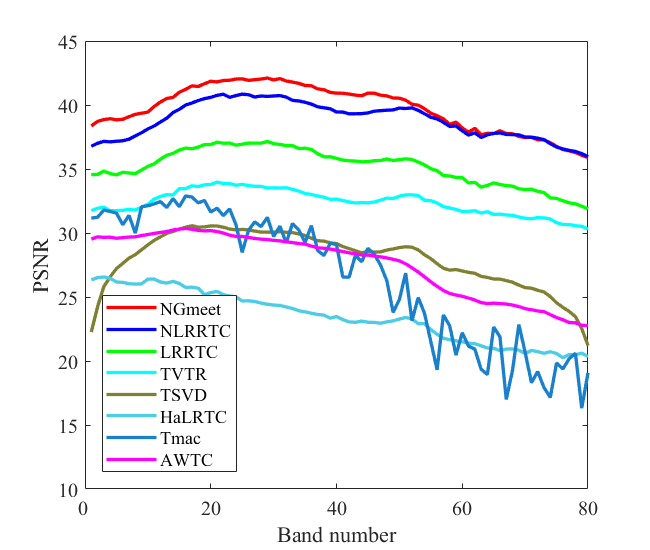}
	}
	\subfigure[SSIM v.s. band]{
		\includegraphics[width= 0.45 \linewidth]{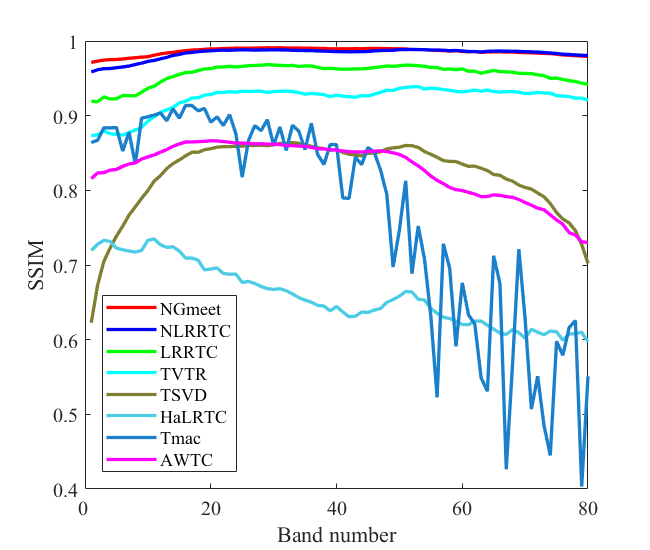}
	}
	%\vspace{-5px}
	\caption{PSNR and SSIM values of each band for different inpainting results on Pavia dataset with SR as $0.05$.}
	%\label{fig:subfig} %% label for entire figure
	\label{fig:inpaint:PSNR}
\end{figure}

\begin{figure*}[ht]
\centering
\includegraphics[width=0.75\linewidth]{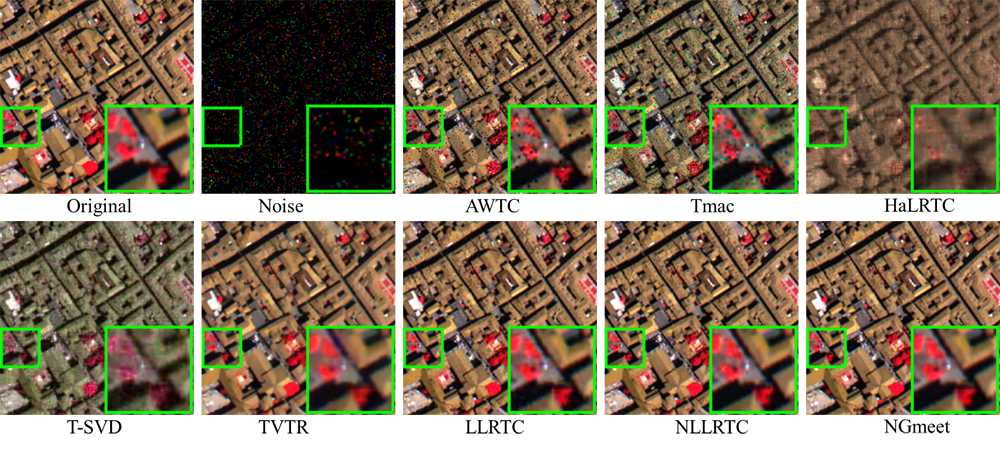}
%\vspace{-10px}
\caption{Inpainting recovered image of different method from PaU dataset with SR as $0.05$. The color image is composed of bands 80, 34, and 9 for the red,
green, and blue channels, respectively.}
\label{fig:inpaint_PaU}
\end{figure*}

\subsubsection{Compressive HSI imaging}

By utilizing the principles of compressed sensing, coded aperture snapshot spectral imagers (CASSI) have been employed for compressive HSI imaging~\cite{arce2013compressive}.
Specifically, wavelength dependent coding is implemented by a coded aperture (physical mask) and a disperser, and the hardware compressed operator is known in advance. So far, GAP-TV~\cite{yuan2016generalized} and DeSCI \cite{Yuan_PAMI_2019}
methods have achieved satisfactory reconstruction results from images coded via CASSI. In this section, we also apply NGmeet to compressive HSI imaging reconstruction from the coded image, with the compressed operator provided by \cite{Yuan_PAMI_2019}. The evaluation data is the CAVE Toy image, which has also been utilized in \cite{Yuan_PAMI_2019}.

The reconstruction results of GAP-TV and DeSCI were provided by \cite{Yuan_PAMI_2019}\footnote{\url{ https://github.com/liuyang12/DeSCI}}.
Tab.~\ref{tab:imaging} presents the evaluation values of NGmeet with those of GAP-TV and DeSCI. Fig.~\ref{fig:img_toy} illustrates different images reconstructed via different methods. It can be concluded that our method NGmeet achieves the best quantitative evaluation results, as well as the best visual results.

\subsection{HSI inpainting experiments}
\label{sec:imp}

%In this section,
%we give the results experiments for HSI with random missing data and remote sensing image stripe inpainting.

\begin{table*}[htbp]
  \centering
  \footnotesize
  \caption{Quantitative comparison of different algorithms in the simulated HSI inpainting experiments.
		The best results are in bold.}
  %\vspace{-10px}
    \begin{tabular}{c|c|c|c|c|c|c|c|c|c|c}
    \hline
    Image & SR   & Index    & AWTC  & Tmac  & HaLRTC & T-SVD & TVTR  & LLRTC & NLLRTC & NGmeet \\ \hline

         &       & PSNR(dB) & 18.02 & 25.27 & 19.81 & 27.90 & 27.93 & 32.80 & 36.80 & \textbf{38.10} \\           \cline{3-11}
     CAVE& 5\%   & SSIM     & 0.500 & 0.746 & 0.645 & 0.781 & 0.753 & 0.896 & 0.978 & \textbf{0.983} \\           \cline{3-11}
         &       & SAM      & 37.573 & 25.602 & 23.068 & 16.304 & 17.517 & 10.279 & 3.498 & \textbf{3.364} \\ \cline{2-11}
         &       & PSNR(dB) & 22.73 & 29.32 & 25.60 & 31.77 & 35.74 & 38.26 & 42.31 & \textbf{44.37} \\ \cline{3-11}
         & 10\%  & SSIM     & 0.653 & 0.799 & 0.798 & 0.882 & 0.944 & 0.965 & 0.992 & \textbf{0.994} \\ \cline{3-11}
         &       & SAM      & 22.725 & 16.452 & 16.656 & 11.781 & 7.529 & 5.916 & 2.202 & \textbf{2.157} \\ \cline{2-11}
         &       & PSNR(dB) & 29.34 & 34.18 & 32.76 & 36.79 & 43.95 & 43.65 & 48.35 & \textbf{49.90} \\ \cline{3-11}
         & 20\%  & SSIM     & 0.807 & 0.928 & 0.934 & 0.952 & 0.991 & 0.988 & \textbf{0.997} & \textbf{0.997} \\ \cline{3-11}
         &       & SAM      & 13.553 & 6.777 & 8.421 & 7.247 & 2.887 & 3.409 & 1.459 & \textbf{1.427} \\ \hline
         &       & PSNR(dB) & 27.53 & 26.62 & 23.47 & 28.03 & 31.86 & 35.19 & 38.84 & \textbf{39.81} \\ \cline{3-11}
     PaU & 5\%   & SSIM     & 0.825 & 0.768 & 0.659 & 0.823 & 0.912 & 0.956 & 0.983 & \textbf{0.986} \\ \cline{3-11}
         &       & SAM      & 7.613 & 12.328 & 9.306 & 10.863 & 5.961 & 5.772 & 3.264 & \textbf{2.974} \\ \cline{2-11}
         &       & PSNR(dB) & 36.92 & 34.69 & 30.08 & 32.01 & 39.03 & 41.16 & 44.79 & \textbf{46.82} \\ \cline{3-11}
         & 10\%  & SSIM  & 0.967 & 0.949 & 0.910 & 0.912 & 0.981 & 0.987 & 0.995 & \textbf{0.996} \\ \cline{3-11}
         &       & SAM   & 3.592 & 3.678 & 6.188 & 8.902 & 3.246 & 3.393 & 2.174 & \textbf{1.792} \\ \cline{2-11}
         &       & PSNR(dB) & 45.99 & 45.55 & 39.89 & 37.00 & 45.69 & 47.91 & 50.00 & \textbf{51.07} \\ \cline{3-11}
         & 20\%  & SSIM  & 0.995 & 0.995 & 0.988 & 0.961 & 0.995 & 0.997 & \textbf{0.998} & \textbf{0.998} \\ \cline{3-11}
         &       & SAM   & 2.026 & 2.003 & 2.932 & 6.500 & 2.004 & 1.824 & 1.455 & \textbf{1.304} \\ \hline
    \end{tabular}%
  \label{tab:inpainting}%
\end{table*}%

\begin{figure*}[ht]
\centering
\includegraphics[width=0.85\linewidth]{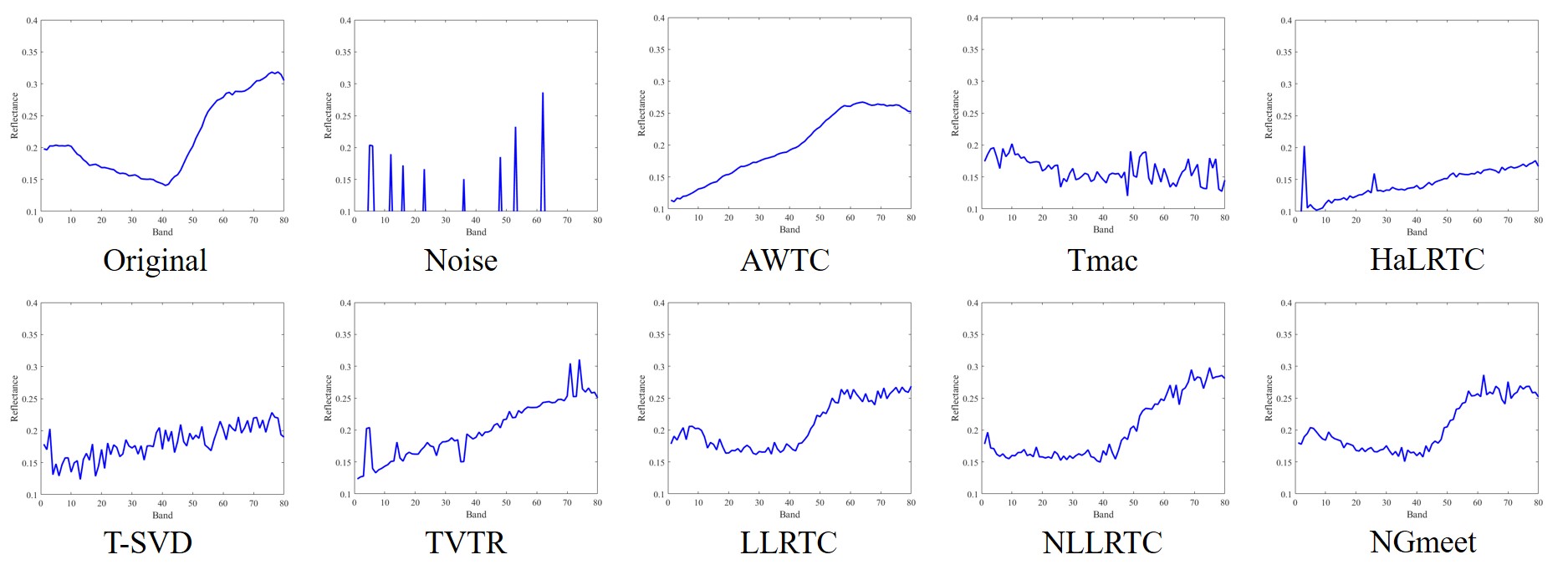}
%\vspace{-15px}
\caption{Spectral signature curves of pixel (50,50) reconstructed by all the compared methods on PaU with SR as $0.05$.}
\label{fig:inpaint:sig}
\end{figure*}

\begin{figure*}[ht]
\centering
\includegraphics[width=0.85\linewidth]{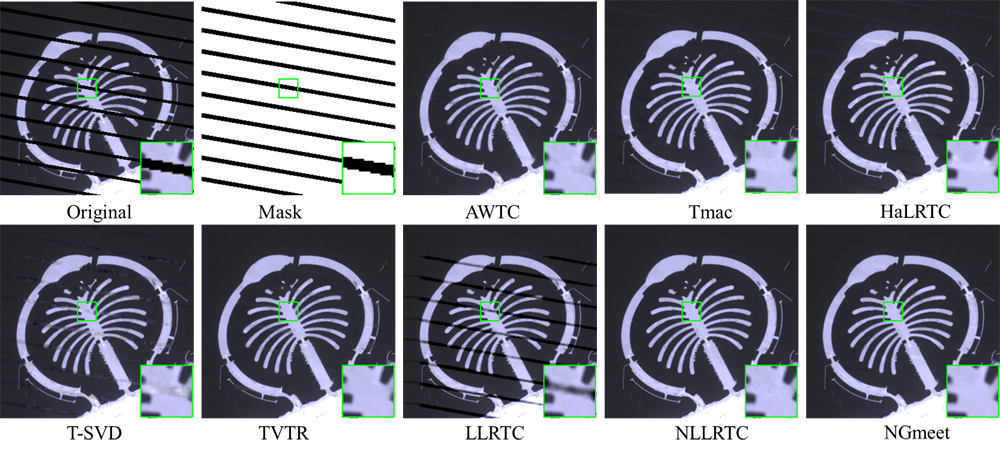}
%\vspace{-15px}
\caption{Reconstructed images of different methods from Landsat-7 dataset.}
\label{fig:inpaint_dubai}
\end{figure*}

\subsubsection{Simulated experiments}

\noindent
\textbf{Setup.}
The MSI CAVE, and HSI PaC images were used~\cite{xie2018tensor}.
Following the settings in~\cite{xie2018tensor}, we selected a subset from CAVE dataset of size $300 \times 300 \times 31$ for the experiments. The PaC image dataset is of size $200 \times 200\times 80$.
The observation ratio of pixels are of $5\%, 10\%$, and $20\%$. Before inpainting, the MSI/HSIs were normalized to [0, 255].
%\footnote{{\color{red}+++ superpixels are randomly missed?--Yes}}

The following methods were compared:
%\footnote{+++ how does hyper-parameters are set for other methods (except ours)? {\color{red} Hyper-parameters of all compared methods are set basedon authors?Ecodes or suggestions in the paper.}}
AWTC~\cite{ng2017adaptive}\footnote{The code was provided by Prof. Qiangqiang Yuan.},
Tmac~\cite{xu2013parallel}\footnote{\url{https://xu-yangyang.github.io/software.html}},
HaLRTC~\cite{liu2012tensor}\footnote{\url{http://www.cs.rochester.edu/u/jliu/}},
T-SVD~\cite{zhang2014novel}\footnote{\url{https://sites.google.com/site/jamiezeminzhang/publications}},
TVTR~\cite{he2019total}\footnote{\url{https://sites.google.com/site/rshewei/home}},
LLRTC~\cite{xie2018tensor},
NLLRTC~\cite{xie2018tensor}\footnote{The code was provided by Dr. Ting Xie.},
%\footnote{\url{http://www.escience.cn/people/changyi/index.html}};
and
NGmeet.
Following~\cite{xie2018tensor}, the proposed NGmeet was initialized as LLRTC.
%Further details can be found in~\cite{xie2018tensor}.

\noindent
\textbf{Quantitative comparison.}
As presented in Tab. \ref{tab:inpainting},
again, NGmeet achieves the best results in almost all cases.
Fig.~\ref{fig:inpaint:PSNR} illustrates the PSNR and SSIM values of each band for different inpainting results on the Pavia dataset with an SR of $5\%$.
NGmeet achieves the highest PSNR and SSIM values almost in all bands, further demonstrating the advantage of the proposed method.

\noindent
\textbf{Visual comparison.}
Fig.~\ref{fig:inpaint_PaU} illustrates the recovered images of different methods from PaU dataset with an SR of $5\%$.
The color image is composed of bands 80, 34, and 9 for the red, green, and blue channels, respectively.
%\footnote{{\color{red}+++ again, better to plot all lines in one figure.Better to plot separately, as the same of related papers}}
Fig.~\ref{fig:inpaint:sig} reports the spectral signature curves of pixel (50,50) reconstructed by all the comparison methods on the PaU dataset. It can be observed that NGmeet always achieves the best visual results.

\subsubsection{Remote sensing stripes inpainting}
Experiments on the real-data stripes inpainting are performed here.
The test data is consist of a time-series Landsat 7 ETM+ image of the Dubai area.
The image size is $300 \times 300 \times 6 \times 8$, which is a spatial size of $300 \times 300$, $6$ spectra, and $8$ time nodes. We merged the spectral and time dimension
and reshaped the image into size $300 \times 300 \times 48$ for the subsequent stripes inpainting.
Fig.~\ref{fig:inpaint_dubai} displays one time node of a remote sensing image before and after inpainting via different methods. The color image is composed of bands 6, 5, and
4. Fig.~\ref{fig:inpaint_dubai}(a-b) show the original missing image and the related masks, respectively.
The visual results NGmeet achieves the best performance.
LRRTC, which is the initialization of NLLRTC and NGmeet, fails to reconstruct the missing image.
However, both NLLRTC and NGmeet can rectify the missing information in LRRTC and achieve the best and second-best inpainting results.

\begin{figure*}[ht]
	\centering
	\subfigure[CAVE toy denoising ($\sigma^2=10$).]
	{\includegraphics[width= 0.25\textwidth]{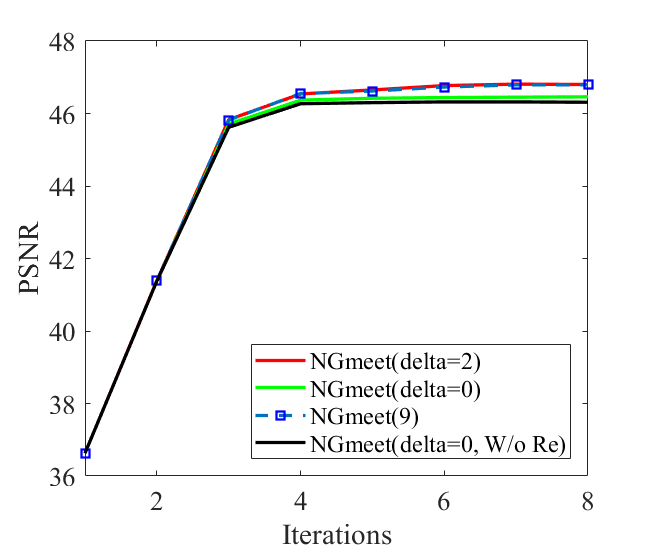}}
	\subfigure[Pavia inpainting ($SR=0.05$).]
	{\includegraphics[width= 0.25\textwidth]{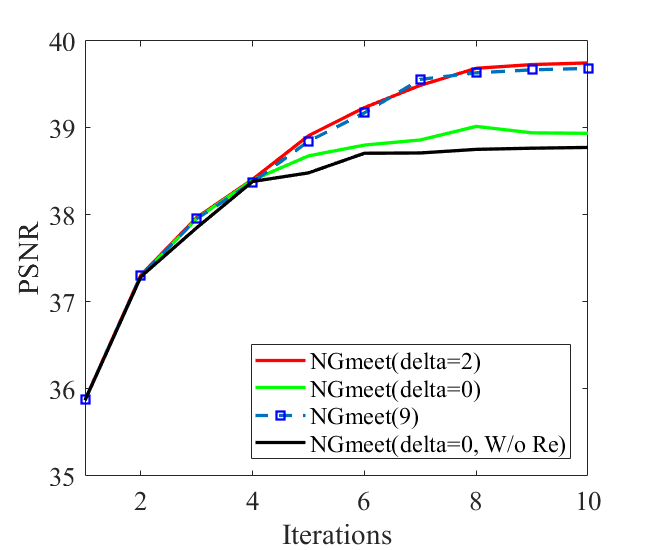}}
	\subfigure[Pavia reconstruction ($SR=0.05$).]
	{\includegraphics[width= 0.25\textwidth]{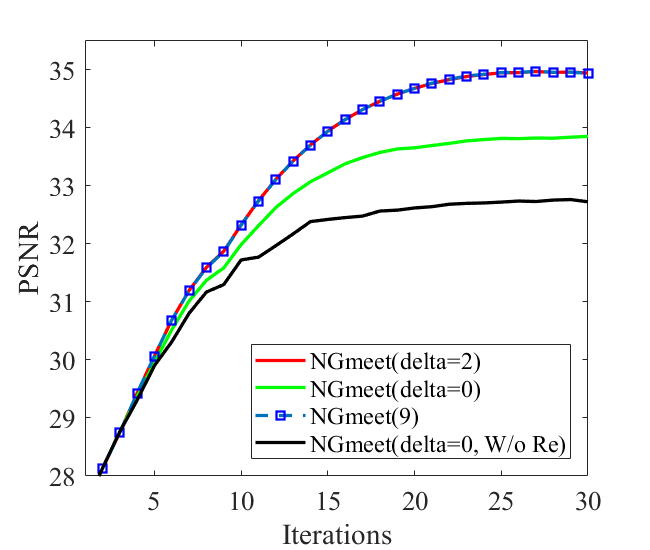}}
	%\vspace{-10px}
	\caption{The comparison between the Algorithm~\ref{alg:NSLR_base} with different empirical settings. NGmeet($\delta=2$) means $\delta=2$ in Algorithm~\ref{alg:NSLR_base}, NGmeet($\delta=0$) means $\delta=0$, NGmeet(9) means adopting \eqref{eq:sub:a_new_r3} to solve \eqref{eq:sub:a_new}, NGmeet($\delta=0$, W/o Re) means $\delta=0$ without re-matching of non-local similar groups.}
	\label{fig:Ite}
\end{figure*}

\subsection{Ablation study}

In this section, we report the results of an ablation study of our NGmeet model.
We adopt the denoising task as the example because HSI restoration shares the same objective model~\eqref{eq:overall_L}
and denoising is the most fundamental task.
%\footnote{{\color{red}+++ I move all experiments which focus on explaining our
%	own method to this section.
%	Please add some sentences to make presentation smooth.
%	e.g., why we only focus on denoising here.}}

\subsubsection{Empirical enhancement analysis}
\label{Empirical:analysis}
We firstly analysis NGmeet (Algorithm~\ref{alg:NSLR_base}) with various empirical enhancement, including empirical solution \eqref{eq:diclearning} and \eqref{eq:sub:a_new_r3} for step 5, re-matching of non-local similar groups and rank adaptation in each iteration. NGmeet($\delta=2$) means $\delta=2$ in Algorithm~\ref{alg:NSLR_base}, NGmeet($\delta=0$) means $\delta=0$, NGmeet(9) means adopting \eqref{eq:sub:a_new_r3} to solve \eqref{eq:sub:a_new}, NGmeet($\delta=0$, W/o Re) means $\delta=0$ without re-matching of non-local similar groups. From Fig.~\ref{fig:Ite}, NGmeet(9) and NGmeet($\delta=2$) perform almost the same in different restoration tasks, indicating the efficiency of optimizing $\mathbf{A}$ via \eqref{eq:diclearning} and \eqref{eq:sub:a_new_r3}, respectively. As the iterations, NGmeet($\delta=2$) can achieve higher PSNR values compared to NGmeet($\delta=0$), further suggesting the advantage of linear updating strategy of $K$. In addition, NGmeet($\delta=0$) can achieve higher PSNR values compared to NGmeet($\delta=0$, W/o Re).
In summary, re-matching of non-local similar groups and rank adaptation of the dimension $K$ in Algorithm~\ref{alg:NSLR_base} disturb the theoretical convergence analysis; however, improve the performance.

\begin{figure}[ht]
	\centering
	\includegraphics[width = 0.5 \linewidth]{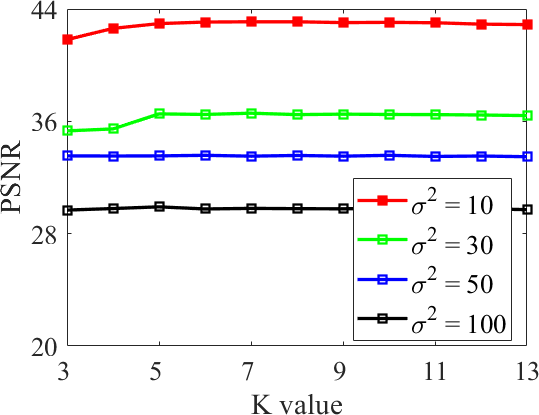}
	%\vspace{-10px}
	\caption{PSNR values achieved by NGmeet with different parameter $K$ with $\delta=0$ on the PaC dataset.}
	\label{fig:K_analysis}
\end{figure}

Subsequently,
we analyze parameter $K$, which is the key to integrating the spatial and spectral information.
The proposed Algorithm \ref{alg:NSLR_base} introduces a linear increase strategy \eqref{delta} to estimate dimension $K$ in each iteration.
The parameter $K$ is controlled by two values, $i.e.,$ the initialization of $K$ and the $stepsize$ $\delta$.
In the experimental section, we adapt HySime~\cite{BioucasTGRS2008} to estimate the initialization of $K$.
The PaC image is chosen as the test image,
and the noise variance $\sigma_0^2$ changes as $10, 30, 50$ and $100$.
Parameter $K$ is initialized by HySime as $7, 6, 6, 5$ for each noise variance cases, respectively.
To further verify the validity of the initialization, we present the PSNR values achieved by NGmeet with different initialization of $K$ with
a $\delta$ of $0$ in Fig.~\ref{fig:K_analysis}. From the figure, it can be observed that the initialization provided by HySime is positive to result in the best results.

As analyzed in Section~\ref{sec:opt},
with the increment of iterations, the noise variance decreases and the parameter $K$ needs to increase to capture more detailed information. We adopt a linearly increase strategy with $stepsize$ $\delta$. Tab.~\ref{tab:delta_analysis} presents the influence of different $\delta$ with different $\sigma_0^2$ values and the related initial $K$ by HySime.
It can be observed that, the updating strategy of $K$ improves the performance. In this paper, we fix the $stepsize$ $\delta$ as $2$ for the whole experiments.

\begin{table}[ht]
	\centering
	\caption{The influence of different $\delta$ for NGmeet.}
	%\vspace{-5px}
	\begin{tabular}{c|c|c|c|c}
		\hline
		PSNR(dB)   & $\sigma_0^2=10$ & $\sigma_0^2=30$ & $\sigma_0^2=50$ & $\sigma_0^2=100$ \\ \hline
		$\delta=0 $  &    43.09    &    36.49    &    33.54    &    29.91     \\ \hline
		$ \delta=1 $ &    43.52    &    36.96    &    34.23    &    30.56     \\ \hline
		$ \delta=2 $ &    43.43    &    37.02    &    34.21    &    30.83     \\ \hline
		$ \delta=3 $ &    43.42    &    37.11    &    34.42    &    30.45     \\ \hline
	\end{tabular}%
	\label{tab:delta_analysis}
\end{table}

\subsubsection{Computational efficiency}

We demonstrate the efficiency of the proposed NGmeet.
%In this section,
%we demonstrate that in our denoising paradigm, the computational efficiency of the non-local denoising procedure will get rid of the huge spectral dimensions.
Compared to the previous non-local denoising methods, \textit{i.e.} KBR~\cite{xie2017kronecker} and LLRT~\cite{chang2017hyper},  NGmeet includes the additional stage A.
%\footnote{+++ please fill in the blanks in the table.{\color{red} filled}}
Tab.~\ref{tab:time} presents the computational time of the different stages of the three methods. These results, along with the complexity analysis in Tab.~\ref{tab:time_com} and~\ref{tab:time} lead us to conclude that
NGmeet takes little time to project the original HSI onto a reduced image (stage A), however, earning huge advantage in stage B including group matching step and non-local denoising.
%\footnote{+++ how do observations here connect with Tab~\ref{tab:time_com}? better to reminder reader the findings in that table as well.}

% Tab generated by Excel2LaTeX from sheet 'Sheet1'
\begin{table}[ht]
	\centering
	\caption{Average running time (in seconds) of each stage for the non-local low-rank based methods.
		stage A: spectral low-rank denoising;
		stage B: spatial non-local low-rank denoising.}
	%\vspace{-10px}
	\footnotesize
	\begin{tabular}{c|c|c|c|c|c}
		\hline
		Time    & KBR  & LLRT &      \multicolumn{3}{c}{NGmeet}       \\ \cline{2-6}
		(seconds) & stage B & stage B & stage A  & stage B & total \\ \hline
		CAVE    & 4330  & 1212  &    3     &  201    &  204 \\ \hline
		PaC    & 828   & 488   &    2     &  37     &  39   \\ \hline
		WDC    & 3570  &1573   &    3     &  45     &  48   \\ \hline
	\end{tabular}
	\label{tab:time}
\end{table}%

\begin{table*}[htbp]
	\centering
	\caption{Quantitative comparison of NGmeet with different non-local low-rank denoising algorithms on the simulated HSI PaU experiments.
		The PSNR is in dB, and best results are in bold.}
	%\vspace{-10px}
	\begin{tabular}{c|c|c|c|c|c|c|c|c|c|c|c|c}
		\hline
		$\sigma$ &             \multicolumn{3}{c|}{10}             &             \multicolumn{3}{c|}{30}             &             \multicolumn{3}{c|}{50}             &             \multicolumn{3}{c}{100}             \\ \hline
		 Index   &      PSNR      &      SSIM      &      SAM      &      PSNR      &      SSIM      &      SAM      &      PSNR      &      SSIM      &      SAM      &      PSNR      &      SSIM      &      SAM      \\ \hline
		  WNNM   &     43.17      &     0.992      &     2.61      &     36.97      &     0.971      &     4.30      &     34.29      &     0.948      &     5.18      &     30.61      &     0.890      &     6.86      \\ \hline
		  WSNM   & \textbf{43.44} & \textbf{0.992} & \textbf{2.45} & \textbf{37.19} & \textbf{0.972} & \textbf{3.95} & \textbf{34.40} & \textbf{0.949} & \textbf{4.74} & \textbf{30.68} & \textbf{0.892} & \textbf{6.66} \\ \hline
	\end{tabular}%
	\label{tab:WNNMvsWSNM}
	%\vspace{-10px}
\end{table*}

\begin{figure}[ht]
	\centering
	\subfigure[Time v.s. number of bands]{\includegraphics[width= 0.45 \linewidth]{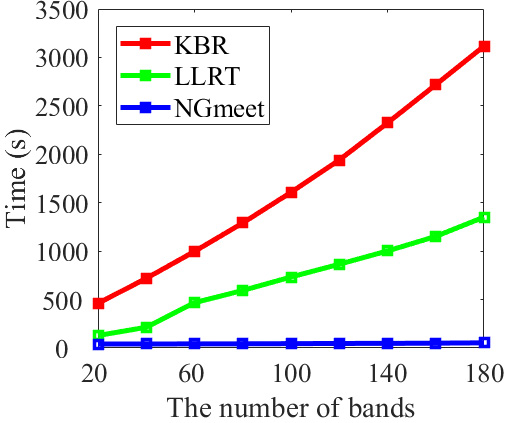}}
	\subfigure[SSIM v.s. number of bands]{\includegraphics[width= 0.45 \linewidth]{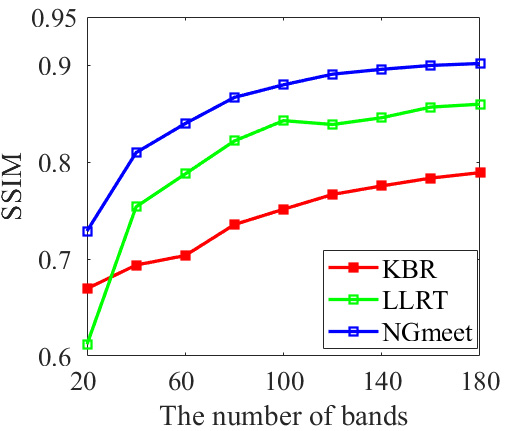}}
	\caption{The computational time and SSIM values of different numbers of bands.
		WDC is used and noise variance is 100.}
	\label{fig:ILL1}
\end{figure}

Fig.~\ref{fig:ILL1} displays the computational time and SSIM values of NGmeet, KBR~\cite{xie2017kronecker} and LLRT~\cite{chang2017hyper}.
As illustrated,
with a larger number of bands,
the computational time significantly increases for KBR and LLRT
but keeps almost the same for NGmeet.
Besides,
the best performance is also consistently best for NGmeet
with different number of bands.

%the computational time also increases linearly.
%Our method can achieve the best performance; moreover, the computational time is nearly unchanged as the number of spectra increases.

\subsubsection{Convergence}
\begin{figure}[ht]
	\centering
	\includegraphics[width=0.5\linewidth]{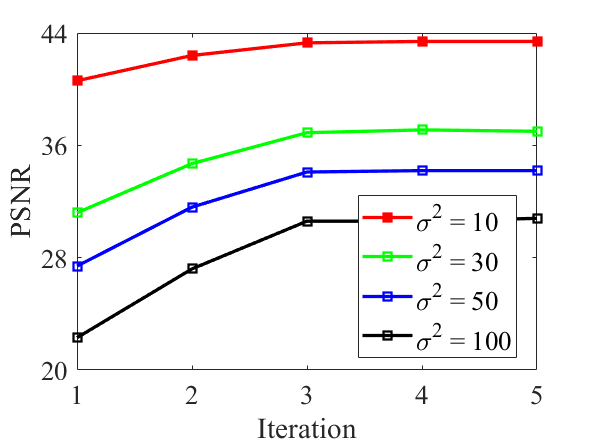}
	\caption{PSNR v.s. iteration of NGmeet. WDC is used.}
	\label{fig:PSNRvsite}
\end{figure}

To show the convergence behavior of NGmeet,
Fig.~\ref{fig:PSNRvsite} presents the PSNR values as the number of iterations increases on the WDC dataset.
The results show that our method can converge to a stable PSNR value very quickly
at different noise levels.

\subsubsection{Choice of denoiser}
\label{sec:denoiser}

In  NGmeet, we choose WNNM~\cite{gu2014weighted} as the denoiser for each non-local patch.
However, WNNM is not the only choice for our proposed NGmeet.
Tab.~\ref{tab:WNNMvsWSNM} reports the results of NGmeet when either WNNM or WSNM~\cite{xieWSNM} is used as the denoiser.
the evaluation results of NGmeet with WSNM are a bit higher than those with WNNM.
Thus, our proposed paradigm is consistent with other advanced denoiser,
and better choice of the reduced image denoiser can further improve the results.

\subsubsection{Comparison with deep learning}

Deep learning has also been introduced to denoise HSIs~\cite{chang2018hsi,yuanHSID-CNN2018}.
In HSI-DeNet~\cite{chang2018hsi},
the model is trained on the ICVL\footnote{\url{https://labicvl.github.io/Datasets_Code.html}} dataset and tested on CAVE dataset.
However, the authors \cite{chang2018hsi} choose only 10 bands (bands 15-24) for training.
For fair comparison, we also implement our proposed NGmeet on CAVE denoising with 10 bands, referred to as NGmeet\_V1. Tab. \ref{tab:DL_com} reports the qualitative evaluation results of HSI-DeNet and NGmeet\_V1 and the original
NGmeet (here, the test results were obtained for all CAVE bands, but only the related 10 bands were evaluated).
From the results, it can be observed that our proposed method outperforms the deep learning based HSI-DeNet method, although HSI-DeNet was trained on the information from the ICVL dataset.
From another side, by comparing NGmeet and NGmeet\_V1, it can be concluded that when more bands are processed simultaneously, better denoising results can be obtained by our method.

\begin{table}[ht]
	\centering
	\caption{Quantitative comparison of between the proposed NGmeet and HSI-DeNet.}
	%\vspace{-10px}
	\begin{tabular}{c|c|c|c|c|c|c}
		\hline
		& \multicolumn{3}{c|}{20} & \multicolumn{3}{c}{50} \\ \hline
		Index    & PSNR  &  SSIM  &  SAM   & PSNR  & SSIM  &  SAM   \\ \hline
		HSI-DeNet  & 38.45 & 0.9741 &  6.41  & 34.59 & 0.892 &  9.51  \\ \hline
		NGmeet\_V1 & 43.03 & 0.9807 &  5.27  & 37.34 & 0.928 &  8.70  \\ \hline
		NGmeet   & 45.01 & 0.9820 &  5.21  & 39.45 & 0.941 &  7.80  \\ \hline
	\end{tabular}%
	\label{tab:DL_com}%
\end{table}%

\section{Conclusion}

In this paper,
we have provided a new perspective about how to integrate spatial non-local similarity and global spectral low-rank property, which are explored using a low-dimensional orthogonal basis and reduced
%\footnote{{\color{red}+++ need revision, no longer just denosing.}}
image denoising, respectively. The unified paradigm was used for the HSI restoration tasks of denoising, compressed HSI reconstruction and inpainting.
We also proposed an alternating minimization method to solve the optimization of the proposed NGmeet method.
The high performance of our method was confirmed by the simulated and real dataset experiments on the three different restoration tasks.
In our unified spatial-spectral paradigm,
the usage of WNNM \cite{gu2014weighted} is not a must.
In future,
we plan to adopt Convolutional Neural Network~\cite{chang2018hsi,zhang2017learning,yuanHSID-CNN2018,dian2018deep} to explore non-local similarity;
and automated machine learning \cite{yao2018taking,yao2020efficient} to help tuning and configuring hyper-parameters. Furthermore, we will also extend our model from Gaussian noise to mixed noise processing.

\section{Acknowledgements}
This work was supported by the Japan Society for the Promotion of Science (KAKENHI 19K20308; KAKENHI 18K18067; KAKENHI 17K00326), the National Key Research and Development Program of China (2018YFB0504500) and the National Natural Science Foundation of China (61871298).

%In our unified spatial-spectral paradigm,
%the usage of WNNM \cite{gu2014weighted} is not a must.
%In future,
%we plan to adopt Convolutional Neural Network~\cite{chang2018hsi,zhang2017learning,yuanHSID-CNN2018} to explore non-local similarity;
%and automated machine learning \cite{yao2018taking} to help tuning and configuring hyper-parameters.

% use section* for acknowledgment
%\ifCLASSOPTIONcompsoc
%  % The Computer Society usually uses the plural form
%  \section*{Acknowledgments}
%\else
%  % regular IEEE prefers the singular form
%  \section*{Acknowledgment}
%\fi
%
%
%The authors would like to thank...

{
%\small
\bibliographystyle{IEEEtr}
\bibliography{lowrank_review_references}

\begin{thebibliography}{10}

\bibitem{Green1998}
R.~O. Green, M.~L. Eastwood, C.~M. Sarture, T.~G. Chrien, M.~Aronsson, B.~J.
  Chippendale, J.~A. Faust, B.~E. Pavri, C.~J. Chovit, M.~Solis, M.~R. Olah,
  and O.~Williams, ``Imaging spectroscopy and the airborne visible/infrared
  imaging spectrometer (aviris),'' {\em Remote Sens. Environ.}, vol.~65,
  pp.~227--248, Sep. 1998.

\bibitem{kwon2005kernel}
H.~Kwon and N.~Nasrabadi, ``Kernel matched subspace detectors for hyperspectral
  target detection,'' {\em IEEE Trans. Pattern Anal. Mach. Intell.}, vol.~28,
  no.~2, pp.~178--194, 2005.

\bibitem{HarvardCVPR2011}
A.~Chakrabarti and T.~Zickler, ``Statistics of real-world hyperspectral
  images,'' in {\em CVPR}, pp.~193--200, 2011.

\bibitem{CAVETIP2010}
F.~Yasuma, T.~Mitsunaga, D.~Iso, and S.~K. Nayar, ``Generalized assorted pixel
  camera: Postcapture control of resolution, dynamic range, and spectrum,''
  {\em IEEE Trans. on Image Process.}, vol.~19, pp.~2241--2253, Sep. 2010.

\bibitem{Stein2002}
D.~Stein, S.~Beaven, L.~Hoff, E.~Winter, A.~Schaum, and A.~Stocker, ``Anomaly
  detection from hyperspectral imagery,'' {\em IEEE Signal Process Mag.},
  vol.~19, no.~1, pp.~58--69, 2002.

\bibitem{Medicalhyper2014}
B.~F. G.~Lu, ``Medical hyperspectral imaging: a review,'' {\em Journal of
  Biomedical Optics}, vol.~19, pp.~19 -- 24, 2014.

\bibitem{facehyperTIP2015}
M.~Uzair, A.~Mahmood, and A.~Mian, ``Hyperspectral face recognition with
  spatiospectral information fusion and pls regression,'' {\em IEEE Trans. on
  Image Process.}, vol.~24, pp.~1127--1137, Mar. 2015.

\bibitem{HyperfaceTPAMI2003}
Z.~Pan, G.~Healey, M.~Prasad, and B.~Tromberg, ``Face recognition in
  hyperspectral images,'' {\em IEEE Trans. Pattern Anal. Mach. Intell.},
  vol.~25, pp.~1552--1560, Dec 2003.

\bibitem{GENDRIN2008}
C.~Gendrin, Y.~Roggo, and C.~Collet, ``Pharmaceutical applications of
  vibrational chemical imaging and chemometrics: A review,'' {\em J. Pharm.
  Biomed. Anal.}, vol.~48, pp.~533 -- 553, Nov. 2008.

\bibitem{chang2017hyper}
Y.~Chang, L.~Yan, and S.~Zhong, ``Hyper-laplacian regularized unidirectional
  low-rank tensor recovery for multispectral image denoising,'' in {\em CVPR},
  pp.~4260--4268, 2017.

\bibitem{Dong2015ICCV}
W.~Dong, G.~Li, G.~Shi, X.~Li, and Y.~Ma, ``Low-rank tensor approximation with
  laplacian scale mixture modeling for multiframe image denoising,'' in {\em
  ICCV}, pp.~442--449, 2015.

\bibitem{CVPR2014Meng}
Y.~Peng, D.~Meng, Z.~Xu, C.~Gao, Y.~Yang, and B.~Zhang, ``Decomposable nonlocal
  tensor dictionary learning for multispectral image denoising,'' in {\em
  CVPR}, pp.~2949--2956, 2014.

\bibitem{xie2017kronecker}
Q.~Xie, Q.~Zhao, D.~Meng, and Z.~Xu, ``Kronecker-basis-representation based
  tensor sparsity and its applications to tensor recovery,'' {\em IEEE Trans.
  Pattern Anal. Mach. Intell.}, vol.~40, no.~8, pp.~1888--1902, 2018.

\bibitem{martin2014hyca}
G.~Martin, J.~Bioucas-Dias, and A.~Plaza, ``Hyca: A new technique for
  hyperspectral compressive sensing,'' {\em IEEE Trans. Geosci. Remote Sens.},
  vol.~53, no.~5, pp.~2819--2831, 2014.

\bibitem{wang2016adaptive}
L.~Wang, Z.~Xiong, G.~Shi, F.~Wu, and W.~Zeng, ``Adaptive nonlocal sparse
  representation for dual-camera compressive hyperspectral imaging,'' {\em IEEE
  Trans. Pattern Anal. Mach. Intell.}, vol.~39, no.~10, pp.~2104--2111, 2016.

\bibitem{zhang2018cluster}
L.~Zhang, W.~Wei, Y.~Zhang, C.~Shen, A.~van, and Q.~Shi, ``Cluster sparsity
  field: An internal hyperspectral imagery prior for reconstruction,'' {\em
  International Journal of Computer Vision}, vol.~126, no.~8, pp.~797--821,
  2018.

\bibitem{he2019total}
W.~{He}, N.~{Yokoya}, L.~{Yuan}, and Q.~{Zhao}, ``Remote sensing image
  reconstruction using tensor ring completion and total variation,'' {\em IEEE
  Trans. Geosci. Remote Sens.}, vol.~57, no.~11, pp.~8998--9009, 2019.

\bibitem{ng2017adaptive}
M.~K. Ng, Q.~Yuan, L.~Yan, and J.~Sun, ``An adaptive weighted tensor completion
  method for the recovery of remote sensing images with missing data,'' {\em
  IEEE Trans. Geosci. Remote Sens.}, vol.~55, no.~6, pp.~3367--3381, 2017.

\bibitem{xie2018tensor}
T.~Xie, S.~Li, L.~Fang, and L.~Liu, ``Tensor completion via nonlocal low-rank
  regularization,'' {\em IEEE trans. on cyber.}, vol.~49, no.~6,
  pp.~2344--2354, 2018.

\bibitem{He2014TGRS}
H.~Zhang, W.~He, L.~Zhang, H.~Shen, and Q.~Yuan, ``Hyperspectral image
  restoration using low-rank matrix recovery,'' {\em IEEE Trans. Geosci. Remote
  Sens.}, vol.~52, pp.~4729--4743, Aug. 2014.

\bibitem{changyiTIP2015}
Y.~Chang, L.~Yan, H.~Fang, and C.~Luo, ``Anisotropic spectral-spatial total
  variation model for multispectral remote sensing image destriping,'' {\em
  IEEE Trans. on Image Process.}, vol.~24, pp.~1852--1866, Jun. 2015.

\bibitem{Zhang_2017_CVPR}
K.~Zhang, W.~Zuo, S.~Gu, and L.~Zhang, ``Learning deep cnn denoiser prior for
  image restoration,'' in {\em CVPR}, July 2017.

\bibitem{yuan2016generalized}
X.~Yuan, ``Generalized alternating projection based total variation
  minimization for compressive sensing,'' in {\em ICIP}, pp.~2539--2543, IEEE,
  Sep. 2016.

\bibitem{baiJSTAR2018}
X.~{Bai}, F.~{Xu}, L.~{Zhou}, Y.~{Xing}, L.~{Bai}, and J.~{Zhou}, ``Nonlocal
  similarity based nonnegative tucker decomposition for hyperspectral image
  denoising,'' {\em IEEE J. Sel.Topics Appl. Earth Observ. Remote Sens.},
  vol.~11, pp.~701--712, March 2018.

\bibitem{DongCS2014}
W.~{Dong}, G.~{Shi}, X.~{Li}, Y.~{Ma}, and F.~{Huang}, ``Compressive sensing
  via nonlocal low-rank regularization,'' {\em IEEE Trans. on Image Process.},
  vol.~23, pp.~3618--3632, Aug 2014.

\bibitem{lishutaoCVPR2017}
R.~Dian, L.~Fang, and S.~Li, ``Hyperspectral image super-resolution via
  non-local sparse tensor factorization,'' in {\em CVPR}, pp.~3862--3871, 2017.

\bibitem{DianTCYB}
R.~{Dian}, S.~{Li}, L.~{Fang}, T.~{Lu}, and J.~M. {Bioucas-Dias}, ``Nonlocal
  sparse tensor factorization for semiblind hyperspectral and multispectral
  image fusion,'' {\em IEEE Trans. on Cyber.},
  vol.~DOI:10.1109/TCYB.2019.2951572, pp.~1--12, 2020.

\bibitem{Bioucas2012jstars}
J.~M. {Bioucas-Dias}, A.~{Plaza}, N.~{Dobigeon}, M.~{Parente}, Q.~{Du},
  P.~{Gader}, and J.~{Chanussot}, ``Hyperspectral unmixing overview:
  Geometrical, statistical, and sparse regression-based approaches,'' {\em IEEE
  J. Sel.Topics Appl. Earth Observ. Remote Sens.}, vol.~5, pp.~354--379, Apr.
  2012.

\bibitem{xie2016hyperspectral}
Y.~Xie, Y.~Qu, D.~Tao, W.~Wu, Q.~Yuan, W.~Zhang, {\em et~al.}, ``Hyperspectral
  image restoration via iteratively regularized weighted schatten p-norm
  minimization.,'' {\em IEEE Trans. Geosci. Remote Sens.}, vol.~54,
  pp.~4642--4659, Aug. 2016.

\bibitem{khan2015joint}
Z.~Khan, F.~Shafait, and A.~Mian, ``Joint group sparse pca for compressed
  hyperspectral imaging,'' {\em IEEE Trans. on Image Process.}, vol.~24,
  no.~12, pp.~4934--4942, 2015.

\bibitem{fan2015exploiting}
B.~Fan, G.~Ely, S.~Aeron, and E.~Miller, ``Exploiting algebraic and structural
  complexity for single snapshot computed tomography hyperspectral imaging
  systems,'' {\em IEEE J. Sel. Topics Signal Process.}, vol.~9, no.~6,
  pp.~990--1002, 2015.

\bibitem{waters2011sparcs}
A.~Waters, A.~C. Sankaranarayanan, and R.~Baraniuk, ``Sparcs: Recovering
  low-rank and sparse matrices from compressive measurements,'' in {\em
  NeurIPS}, pp.~1089--1097, 2011.

\bibitem{xu2013parallel}
Y.~Xu, R.~Hao, W.~Yin, and Z.~Su, ``Parallel matrix factorization for low-rank
  tensor completion,'' {\em arXiv preprint arXiv:1312.1254}, 2013.

\bibitem{zhuang2018fast}
L.~Zhuang and J.~Bioucas-Dias, ``Fast hyperspectral image denoising and
  inpainting based on low-rank and sparse representations,'' {\em IEEE J.
  Sel.Topics Appl. Earth Observ. Remote Sens.}, vol.~11, pp.~730--742, Mar.
  2018.

\bibitem{yao2019efficient}
Q.~Yao, J.~Kwok, and B.~Han, ``Efficient nonconvex regularized tensor
  completion with structure-aware proximal iterations,'' in {\em ICML},
  pp.~7035--7044, 2019.

\bibitem{HE2016TGRS}
W.~He, H.~Zhang, L.~Zhang, and H.~Shen, ``Total-variation-regularized low-rank
  matrix factorization for hyperspectral image restoration,'' {\em IEEE Trans.
  Geosci. Remote Sens.}, vol.~54, pp.~178--188, Jan. 2016.

\bibitem{lu2013graph}
X.~Lu, Y.~Wang, and Y.~Yuan, ``Graph-regularized low-rank representation for
  destriping of hyperspectral images,'' {\em IEEE Trans. Geosci. Remote Sens.},
  vol.~51, no.~7, pp.~4009--4018, 2013.

\bibitem{fu2017adaptive}
Y.~Fu, A.~Lam, I.~Sato, and Y.~Sato, ``Adaptive spatial-spectral dictionary
  learning for hyperspectral image restoration,'' {\em IJCV}, vol.~122, no.~2,
  pp.~228--245, 2017.

\bibitem{golbabaee2012joint}
M.~Golbabaee and P.~Vandergheynst, ``Joint trace/tv norm minimization: A new
  efficient approach for spectral compressive imaging,'' in {\em ICIP},
  pp.~933--936, IEEE, 2012.

\bibitem{peng2018enhanced}
J.~Peng, Q.~Xie, Q.~Zhao, Y.~Wang, D.~Meng, and Y.~Leung, ``Enhanced 3dtv
  regularization and its applications on hyper-spectral image denoising and
  compressed sensing,'' {\em arXiv preprint arXiv:1809.06591}, 2018.

\bibitem{wang2017compressive}
Y.~Wang, L.~Lin, Q.~Zhao, T.~Yue, D.~Meng, and Y.~Leung, ``Compressive sensing
  of hyperspectral images via joint tensor tucker decomposition and weighted
  total variation regularization,'' {\em IEEE Geos. and Remote Sens.Lett.},
  vol.~14, no.~12, pp.~2457--2461, 2017.

\bibitem{rasti2014hyperspectral}
B.~Rasti, J.~Sveinsson, M.~Ulfarsson, and J.~Benediktsson, ``Hyperspectral
  image denoising using first order spectral roughness penalty in wavelet
  domain,'' {\em IEEE J. Sel. Topics Appl. Earth Observ. Remote Sens}, vol.~7,
  pp.~2458--2467, Jun. 2014.

\bibitem{dian2019hyperspectral}
R.~{Dian} and S.~{Li}, ``Hyperspectral image super-resolution via
  subspace-based low tensor multi-rank regularization,'' {\em IEEE Trans. on
  Image Process.}, vol.~28, no.~10, pp.~5135--5146, 2019.

\bibitem{dian2019learning}
R.~{Dian}, S.~{Li}, and L.~{Fang}, ``Learning a low tensor-train rank
  representation for hyperspectral image super-resolution,'' {\em IEEE Trans.
  Neural Netw. Learn. Syst.}, vol.~30, no.~9, pp.~2672--2683, 2019.

\bibitem{FuCVPR2016}
Y.~Fu, Y.~Zheng, I.~Sato, and Y.~Sato, ``Exploiting spectral-spatial
  correlation for coded hyperspectral image restoration,'' in {\em CVPR},
  pp.~3727--3736, June 2016.

\bibitem{MengTIP2016}
X.~Cao, Q.~Zhao, D.~Meng, Y.~Chen, and Z.~Xu, ``Robust low-rank matrix
  factorization under general mixture noise distributions,'' {\em IEEE Trans.
  on Image Process.}, vol.~25, pp.~4677--4690, Oct. 2016.

\bibitem{weiwei2015CVPR}
L.~Zhang, W.~Wei, Y.~Zhang, C.~Tian, and F.~Li, ``Reweighted laplace prior
  based hyperspectral compressive sensing for unknown sparsity,'' in {\em
  CVPR}, Jun. 2015.

\bibitem{he2018non}
W.~He, Q.~Yao, C.~Li, N.~Yokoya, and Q.~Zhao, ``Non-local meets global: An
  integrated paradigm for hyperspectral denoising,'' in {\em CVPR}, Jun. 2019.

\bibitem{Kolda2009}
T.~Kolda and B.~Bader, ``Tensor decompositions and applications,'' {\em SIAM
  Review}, vol.~51, no.~3, pp.~455--500, 2009.

\bibitem{xue2019nonlocal}
J.~Xue, Y.~Zhao, W.~Liao, and J.~C. Chan, ``Nonlocal tensor sparse
  representation and low-rank regularization for hyperspectral image
  compressive sensing reconstruction,'' {\em Remote Sens.}, vol.~11, no.~2,
  p.~193, 2019.

\bibitem{Yuan_PAMI_2019}
Y.~{Liu}, X.~{Yuan}, J.~{Suo}, D.~J. {Brady}, and Q.~{Dai}, ``Rank minimization
  for snapshot compressive imaging,'' {\em IEEE Trans. Pattern Anal. Mach.
  Intell.}, vol.~41, pp.~2990--3006, Dec 2019.

\bibitem{zhang2019computational}
S.~Zhang, L.~Wang, Y.~Fu, X.~Zhong, and H.~Huang, ``Computational hyperspectral
  imaging based on dimension-discriminative low-rank tensor recovery,'' in {\em
  ICCV}, pp.~10183--10192, 2019.

\bibitem{chang2017weighted}
Y.~{Chang}, L.~{Yan}, X.~{Zhao}, H.~{Fang}, Z.~{Zhang}, and S.~{Zhong},
  ``Weighted low-rank tensor recovery for hyperspectral image restoration,''
  {\em IEEE Trans. on Cyber.}, pp.~1--15, 2020.

\bibitem{Yong2020TIP}
Y.~{Chen}, T.~{Huang}, W.~{He}, N.~{Yokoya}, and X.~{Zhao}, ``Hyperspectral
  image compressive sensing reconstruction using subspace-based nonlocal tensor
  ring decomposition,'' {\em IEEE Trans. on Image Process.},
  vol.~DOI:10.1109/TIP.2020.2994411, pp.~1--1, 2020.

\bibitem{ji2018nonlocal}
T.~Ji, N.~Yokoya, X.~Zhu, and T.~Huang, ``Nonlocal tensor completion for
  multitemporal remotely sensed images inpainting,'' {\em IEEE Trans. Geosci.
  Remote Sens.}, vol.~56, no.~6, pp.~3047--3061, 2018.

\bibitem{BioucasTGRS2008}
J.~M. {Bioucas-Dias} and J.~M.~P. {Nascimento}, ``Hyperspectral subspace
  identification,'' {\em IEEE Trans. Geosci. Remote Sens.}, vol.~46,
  pp.~2435--2445, Aug. 2008.

\bibitem{arce2013compressive}
G.~Arce, D.~Brady, L.~Carin, H.~Arguello, and D.~Kittle, ``Compressive coded
  aperture spectral imaging: An introduction,'' {\em IEEE Signal Process Mag.},
  vol.~31, no.~1, pp.~105--115, 2013.

\bibitem{fowler2009compressive}
J.~Fowler, ``Compressive-projection principal component analysis,'' {\em IEEE
  trans. on image process.}, vol.~18, pp.~2230--2242, Jun. 2009.

\bibitem{renard2008denoising}
N.~{Renard}, S.~{Bourennane}, and J.~{Blanc-Talon}, ``Denoising and
  dimensionality reduction using multilinear tools for hyperspectral images,''
  {\em IEEE Geosci. Remote Sens. Lett.}, vol.~5, pp.~138--142, April 2008.

\bibitem{gong2020low}
X.~Gong, W.~Chen, and J.~Chen, ``A low-rank tensor dictionary learning method
  for hyperspectral image denoising,'' {\em IEEE Trans. on Image Process.},
  vol.~68, pp.~1168--1180, 2020.

\bibitem{chen2019hyperspectral}
Y.~Chen, W.~He, N.~Yokoya, and T.-Z. Huang, ``Hyperspectral image restoration
  using weighted group sparsity-regularized low-rank tensor decomposition,''
  {\em IEEE trans. on cyber.}, 2019.

\bibitem{cao2016total}
W.~Cao, Y.~Wang, J.~Sun, D.~Meng, C.~Yang, A.~Cichocki, and Z.~Xu, ``Total
  variation regularized tensor rpca for background subtraction from compressive
  measurements,'' {\em IEEE Trans. on Image Process.}, vol.~25, no.~9,
  pp.~4075--4090, 2016.

\bibitem{ji2016tensor}
T.~Ji, T.~Huang, X.~Zhao, T.~Ma, and G.~Liu, ``Tensor completion using total
  variation and low-rank matrix factorization,'' {\em Information Sciences},
  vol.~326, pp.~243--257, 2016.

\bibitem{zhuang2017hyperspectral}
L.~Zhuang and J.~Bioucas-Dias, ``Hyperspectral image denoising based on global
  and non-local low-rank factorizations,'' in {\em ICIP}, pp.~1900--1904, IEEE,
  2017.

\bibitem{cao2019hyperspectral}
C.~Cao, J.~Yu, C.~Zhou, K.~Hu, F.~Xiao, and X.~Gao, ``Hyperspectral image
  denoising via subspace-based nonlocal low-rank and sparse factorization,''
  {\em IEEE J. Sel.Topics Appl. Earth Observ. Remote Sens.}, vol.~12, no.~3,
  pp.~973--988, 2019.

\bibitem{zhang2017learning}
K.~Zhang, W.~Zuo, S.~Gu, and L.~Zhang, ``Learning deep cnn denoiser prior for
  image restoration,'' in {\em CVPR}, vol.~2, 2017.

\bibitem{chang2018hsi}
Y.~Chang, L.~Yan, H.~Fang, S.~Zhong, and W.~Liao, ``Hsi-denet: Hyperspectral
  image restoration via convolutional neural network,'' {\em IEEE Trans.
  Geosci. Remote Sens.}, pp.~1--16, 2018.

\bibitem{DianTNNLS}
R.~{Dian}, S.~{Li}, and X.~{Kang}, ``Regularizing hyperspectral and
  multispectral image fusion by cnn denoiser,'' {\em IEEE Trans. Neural Netw.
  Learn. Syst.}, vol.~DOI:10.1109/TNNLS.2020.2980398, pp.~1--12, 2020.

\bibitem{nocedal2006numerical}
J.~Nocedal and S.~Wright, {\em Numerical optimization}.
\newblock Springer, 2006.

\bibitem{wang2008new}
Y.~Wang, J.~Yang, W.~Yin, and Y.~Zhang, ``A new alternating minimization
  algorithm for total variation image reconstruction,'' {\em SIAM Journal on
  Imaging Sciences}, vol.~1, no.~3, pp.~248--272, 2008.

\bibitem{rudin1992nonlinear}
L.~Rudin, S.~Osher, and E.~Fatemi, ``Nonlinear total variation based noise
  removal algorithms,'' {\em Physica D: nonlinear phenomena}, vol.~60, no.~1-4,
  pp.~259--268, 1992.

\bibitem{mairal2009non}
J.~Mairal, F.~Bach, J.~Ponce, G.~Sapiro, and A.~Zisserman, ``Non-local sparse
  models for image restoration.,'' in {\em ICCV}, vol.~29, pp.~54--62,
  Citeseer, 2009.

\bibitem{dong2013nonlocal}
W.~Dong, G.~Shi, and X.~Li, ``Nonlocal image restoration with bilateral
  variance estimation: a low-rank approach,'' {\em IEEE Trans. on Image
  Process.}, vol.~22, no.~2, pp.~700--711, 2013.

\bibitem{gu2014weighted}
S.~Gu, L.~Zhang, W.~Zuo, and X.~Feng, ``Weighted nuclear norm minimization with
  application to image denoising,'' in {\em CVPR}, pp.~2862--2869, 2014.

\bibitem{parikh2014proximal}
N.~Parikh, S.~Boyd, {\em et~al.}, ``Proximal algorithms,'' {\em Foundations and
  Trends{\textregistered} in Optimization}, vol.~1, no.~3, pp.~127--239, 2014.

\bibitem{bauschke2008proximal}
H.~H. Bauschke, R.~Goebel, Y.~Lucet, and X.~Wang, ``The proximal average: basic
  theory,'' {\em SIAM Journal on Optimization}, vol.~19, no.~2, pp.~766--785,
  2008.

\bibitem{yu2013better}
Y.-L. Yu, ``Better approximation and faster algorithm using the proximal
  average,'' in {\em NeurIPS}, pp.~458--466, 2013.

\bibitem{tao2005dc}
P.~D. Tao, ``The {DC} (difference of convex functions) programming and {DCA}
  revisited with {DC} models of real world nonconvex optimization problems,''
  {\em Annals of Operations Research}, vol.~133, no.~1-4, pp.~23--46, 2005.

\bibitem{he2015hyperspectral}
W.~He, H.~Zhang, L.~Zhang, and H.~Shen, ``Hyperspectral image denoising via
  noise-adjusted iterative low-rank matrix approximation,'' {\em IEEE J.
  Sel.Topics Appl. Earth Observ. Remote Sens.}, vol.~8, no.~6, pp.~3050--3061,
  2015.

\bibitem{QianyuntaoTGRS2015}
M.~Ye, Y.~Qian, and J.~Zhou, ``Multitask sparse nonnegative matrix
  factorization for joint spectral-spatial hyperspectral imagery denoising,''
  {\em IEEE Trans. Geosci. Remote Sens.}, vol.~53, pp.~2621--2639, May 2015.

\bibitem{liu2012denoising}
X.~Liu, S.~Bourennane, and C.~Fossati, ``Denoising of hyperspectral images
  using the parafac model and statistical performance analysis,'' {\em IEEE
  Trans. Geosci. Remote Sens.}, vol.~50, no.~10, pp.~3717--3724, 2012.

\bibitem{SSIM2004image}
Z.~Wang, A.~C. Bovik, H.~R. Sheikh, and E.~P. Simoncelli, ``Image quality
  assessment: from error visibility to structural similarity,'' {\em IEEE
  Trans. on Image Process.}, vol.~13, no.~4, pp.~600--612, 2004.

\bibitem{daubechies2004iterative}
I.~Daubechies, M.~Defrise, and C.~Mol, ``An iterative thresholding algorithm
  for linear inverse problems with a sparsity constraint,'' {\em Communications
  on Pure and Applied Mathematics: A Journal Issued by the Courant Institute of
  Mathematical Sciences}, vol.~57, no.~11, pp.~1413--1457, 2004.

\bibitem{martin2016hyperspectral}
G.~Mart{\'\i}n and J.~M. Bioucas-Dias, ``Hyperspectral blind reconstruction
  from random spectral projections,'' {\em IEEE J. Sel.Topics Appl. Earth
  Observ. Remote Sens.}, vol.~9, pp.~2390--2399, Jun. 2016.

\bibitem{zhang2016dictionary}
L.~Zhang, W.~Wei, Y.~Zhang, C.~Shen, A.~Van, and Q.~Shi, ``Dictionary learning
  for promoting structured sparsity in hyperspectral compressive sensing,''
  {\em IEEE Trans. Geosci. Remote Sens.}, vol.~54, pp.~7223--7235, Dec. 2016.

\bibitem{liu2012tensor}
J.~Liu, P.~Musialski, P.~Wonka, and J.~Ye, ``Tensor completion for estimating
  missing values in visual data,'' {\em IEEE Trans. Pattern Anal. Mach.
  Intell.}, vol.~35, pp.~208--220, Jan. 2012.

\bibitem{zhang2014novel}
Z.~Zhang, G.~Ely, S.~Aeron, N.~Hao, and M.~Kilmer, ``Novel methods for
  multilinear data completion and de-noising based on tensor-svd,'' in {\em
  CVPR}, pp.~3842--3849, 2014.

\bibitem{xieWSNM}
Y.~{Xie}, S.~{Gu}, Y.~{Liu}, W.~{Zuo}, W.~{Zhang}, and L.~{Zhang}, ``Weighted
  schatten$p$-norm minimization for image denoising and background
  subtraction,'' {\em IEEE Trans. on Image Process.}, vol.~25, pp.~4842--4857,
  Oct 2016.

\bibitem{yuanHSID-CNN2018}
Q.~Yuan, Q.~Zhang, J.~Li, H.~Shen, and L.~Zhang, ``Hyperspectral image
  denoising employing a spatial-spectral deep residual convolutional neural
  network,'' {\em IEEE Trans. Geosci. Remote Sens.}, pp.~1--14, 2018.

\bibitem{dian2018deep}
R.~Dian, S.~Li, A.~Guo, and L.~Fang, ``Deep hyperspectral image sharpening,''
  {\em IEEE Trans. Neural Netw.}, vol.~29, no.~11, pp.~5345--5355, 2018.

\bibitem{yao2018taking}
Q.~Yao, M.~Wang, Y.~Chen, W.~Dai, H.~Y., Y.~Li, W.~Tu, Q.~Yang, and Y.~Yu,
  ``Taking human out of learning applications: A survey on automated machine
  learning,'' tech. rep., arXiv preprint, 2018.

\bibitem{yao2020efficient}
Q.~Yao, J.~Xu, W.-W. Tu, and Z.~Zhu, ``Efficient neural architecture search via
  proximal iterations.,'' in {\em AAAI}, pp.~6664--6671, 2020.

\end{thebibliography}
}

\end{document}